\documentclass[11pt]{article}
\usepackage[T1]{fontenc}
\usepackage[latin9]{inputenc}
\usepackage{geometry}
\usepackage{amsmath}
\usepackage{amssymb}
\usepackage{graphicx}

\makeatletter

\providecommand{\tabularnewline}{\\}
\usepackage{tikz}
\usetikzlibrary{calc}

\newcommand{\lyxaddress}[1]{
	\par {\raggedright #1
	\vspace{1.4em}
	\noindent\par}
}

\makeatletter 
\g@addto@macro\bfseries{\boldmath}
\makeatother

\usepackage[labelfont=bf, format=hang]{caption}

\usepackage{mathrsfs}

\newcommand\fverb{\setbox\fverbbox=\hbox\bgroup\verb}
\newcommand\fverbdo{\egroup\medskip\noindent%
			\fbox{\unhbox\fverbbox}\ }
\newcommand\fverbit{\egroup\item[\fbox{\unhbox\fverbbox}]}
\newbox\fverbbox

\newcommand{\pslash}{p\kern-1ex /}
\newcommand{\qslash}{q\kern-1ex /}
\newcommand{\lslash}{l\kern-1ex /}
\newcommand{\sslash}{s\kern-1ex /}
\newcommand{\kaslash}{k_a\kern-2ex /}
\newcommand{\kbslash}{k_b\kern-2ex /}
\newcommand{\Dslash}{\mathcal{D}\kern-1.5ex /}

\newcommand{\beqa}{\begin{eqnarray}}
\newcommand{\eeqa}{\end{eqnarray}}

\newcommand{\ba}{\begin{eqnarray}}
\newcommand{\ea}{\end{eqnarray}}
\newcommand{\be}{\begin{equation}}

\newcommand{\beq}{\begin{eqnarray}}
\newcommand{\eeq}{\end{eqnarray}}

\newcommand{\ee}{\end{equation}}

\newcommand{\bm}{\begin{multline}}
\newcommand{\fm}{\end{multline}}

\makeatother

\begin{document}
\title{{\huge{}Resurgence in the $O(4)$ sigma model}}
\author{\hspace{-7mm}Michael C. Abbott$^{1}$, Zoltán Bajnok$^{1}$, János
Balog$^{1}$, Árpád Heged\H{u}s$^{1}$ \& Saeedeh Sadeghian$^{1,2}$}
\date{{\normalsize{}\medskip v1: 24 November 2020}}
\maketitle

\lyxaddress{\begin{center}
$^{1}$\emph{Wigner Research Centre for Physics,}\\
\emph{Konkoly-Thege Miklós u. 29-33, 1121 Budapest, Hungary\medskip }\\
$^{2}$\emph{Department of Theoretical Physics, Faculty of Basic Sciences,
}\\
\emph{University of Mazandaran, P.O. Box 47416-95447, Babolsar, Iran}
\par\end{center}}
\begin{abstract}
\noindent We analyze the free energy of the integrable two dimensional
O(4) sigma model in a magnetic field. We use Volin's method to extract
high number (2000) of perturbative coefficients with very high precision.
The factorial growth of these coefficients are regulated by switching
to the Borel transform, where we perform several asymptotic analysis.
High precision data allowed to identify Stokes constants and alien
derivatives with exact expressions. These reveal a nice resurgence
structure which enables to formulate the first few terms of the ambiguity
free trans-series. We check these results against the direct numerical
solution of the exact integral equation and find complete agreement.
\end{abstract}
\tableofcontents{}

\section{Introduction}

Perturbation theory has proved to be a useful tool in calculating
physical processes for the electromagnetic and weak interactions,
but has had only a limited success for their strong counterpart. Important
phenomena such as confinement and dynamical mass generation are inherently
non-perturbative, and cannot be accessed from the few known perturbative
coefficients. Perturbation theory in QCD is expected to be asymptotic,
with coefficients growing factorially, see e.g \cite{Bauer:2011ws,Caprini:2020lff}.
This factorial growth can be traced to the proliferation of Feynman
diagrams \cite{Hurst:1952zh,Lipatov:1976ny}, or to integrals in specific
renormalon diagrams, when loop momenta lie in various IR and UV domains
\cite{Beneke:1998ui}. It is a signal of non-perturbative contributions,
which usually originate from non-trivial saddle points in the path
integral. Because these are exponentially suppressed, they do not
appear directly in perturbation theory, but they can be extracted
from the large-order behaviour of perturbative coefficients. The tools
for doing this are known as resurgence theory \cite{Dorigoni:2014hea,Dunne:2015eaa,Aniceto:2018bis}.

While it would be ideal to apply this theory to non-perturbative phenomena
in QCD, too few perturbative coefficients are known for this to make
progress. Thus we turn to toy models, which share important features
with QCD, but are more tractable. The two dimensional $O(N)$ symmetric
sigma models are exactly of this type: They exhibit a dynamically
generated mass gap, and are asymptotically free in perturbation theory
\cite{Hasenfratz:1990zz}. At the same time they are integrable, allowing
physical quantities such as the mass gap, scattering matrices, and
ground state energy to be calculated exactly \cite{Zamolodchikov:1977nu,Hasenfratz:1990zz}.
Our aim in the present paper is to use the $O(4)$ sigma model to
reveal the relation between the perturbative and non-perturbative
effects, as the first steps in the full resurgence program.

The simplest cure for the factorial growth of perturbative coefficients
is the Borel transform, which is obtained by dividing out this factorial
in order to ensure constant asymptotics. The resulting function has
a finite radius of convergence, and exhibits pole and cut singularities,
often on the real line. Since the inverse Borel (Laplace) transformation
involves an integration of the analytically continued function along
the positive real line, singularities there lead to ambiguities in
the result, as the contour must be shifted. Such cases are often called
non Borel summable, but in fact these ambiguities, which are imaginary
and exponentially small, contain useful information. While isolated
poles give single exponential terms, each cut gives an exponential
multiplied by a power series, which is itself asymptotic. The coefficients
of the original series are said to resurge into those of these further
series, which describe the contribution of non-trivial non-perturbative
saddle points. The maps between these different sectors are called
alien derivatives, and their algebraic properties provide useful constraints.
The full physical expression is what is called a trans-series, a sum
over different exponential factors each multiplying a real-valued
power series, which is the final result.

The resurgence program has been pushed forward for several quantum
field theories including supersymmetric theories \cite{Aniceto:2011nu,Aniceto:2014hoa,Fujimori:2016ljw,Dorigoni:2017smz,Fujimori:2018kqp,Dorigoni:2019kux},
various quantities in the maximally supersymmetric 4D gauge theory
in the large $N$ limit \cite{Aniceto:2015rua,Dorigoni:2015dha,Arutyunov:2016etw},
the simple $\phi^{4}$ theory in 2D, \cite{Serone:2018gjo,Serone:2019szm}.
Factorial growth can be seen in lattice simulations \cite{Bruckmann:2019mky}.
Much is known in the large $N$, \cite{Dunne:2016nmc,Marino:2012zq}
and also semiclassically \cite{Cherman:2013yfa,Misumi:2014bsa,Dunne:2015ywa}.
They also appear in the hydrodynamics of the Yang Mills plasma \cite{Aniceto:2015mto,Aniceto:2018uik}.
Recently in many body systems the authors of \cite{Marino:2019fuy,Marino:2019wra}
could manage to extract the leading exponentially small corrections
exactly in the Gaudin-Yang model. In the Hubbard model the knowledge
of the exact perturbative coefficients enabled them to calculate all
the alien derivatives \cite{Marino:2020dgc}. We are not aware of
any asymptotically free QFT, however, where the full resurgence theory
had been rigorously established.

The $O(4)$ sigma model could be the first example, as it allows an
exact treatment. This model was one of the first where the scattering
matrix was exactly determined \cite{Zamolodchikov:1977nu}. It was
also the first model where the dynamically generated scale $\Lambda_{\overline{MS}}$
was analytically related to the mass of the particles $m$ \cite{Hasenfratz:1990zz}.
This seminal calculation was done by introducing a magnetic field
coupled to one of the conserved $O(N)$ charges, and determining the
free energy in two different ways. For large magnetic fields one can
establish a standard, renormalization group improved perturbative
expansion in the parameter $h/\Lambda_{\overline{MS}}$. On the other
hand, the magnetic field forces the positively charged particles to
condense into the vacuum, and this vacuum condensate consists of particles
with rapidities in $(-B,B)$, with density $\chi(\theta)$. Based
on the scattering matrix the thermodynamic limit of the Bethe ansatz
leads to a linear integral (TBA) equation for $\chi(\theta)$, which
determines the density $\rho$ and the groundstate energy $\epsilon$,
whose Legendre transform is the sought for free energy. By neglecting
exponentially small contributions the TBA equation can be expanded
as a function of $h/m$. Comparison to the perturbative expansion
then led to the relation between the UV parameter $\Lambda_{\overline{MS}}$
and the IR parameter $m$. We emphasize that the TBA equation is exact,
in that it contains also all the non-perturbative exponentially small
corrections. Thus its analytical expansion could lead directly to
the exact trans-series of the groundstate energy and free energy,
which would provide a veritable gold mine for the resurgence literature.
Unfortunately, the exact calculation of the exponentially small terms
is beyond the scope of the present day research. Even the calculation
of not just the first few perturbative coefficients resisted an analytical
treatment for decades.

A breakthrough was obtained by Volin, who invented a way to calculate
the perturbative coefficients systematically \cite{Volin:2009wr,Volin:2010cq}.
His idea was to expand the resolvent of the spectral density $\chi(\theta)$
both in the middle of the interval $\theta\sim0$ and in the edge
region $\theta\sim B$, and then to match the two expansions. In the
middle region the TBA equation determines the analytical structure
of the resolvent, while in the edge region the Wiener-Hopf technique
can be used to parameterize its Laplace transform. Surprisingly, matching
the two representations fixes all the perturbative coefficients in
terms of algebraic equations, which can be solved iteratively. Using
this method Volin calculated the first $26$ perturbative coefficients
for generic $O(N)$ models, which was later extended to $44$ coefficients
in \cite{Marino:2019eym}. These authors also extended the method
for other relativistic and non-relativistic theories \cite{Marino:2019fuy,Marino:2019wra}.
The perturbative coefficients in the $O(N)$ models are linear combinations
of products of odd zeta functions degrees not larger than the perturbative
order. Although this information allowed the authors to gain qualitative
information about the analytical behavior of the Borel transform,
and establish the location of the leading singularities \cite{Volin:2009wr,Marino:2019eym},
it is not sufficient to investigate resurgence properties. Focusing
on the $O(4)$ model, we managed to solve the algebraic equations
recursively in a closed form. This enabled us to calculate the first
50 perturbative coefficients exactly, and switching to work numerically,
the first $2000$ terms with $12000$ digits precision. Numerical
data is equally useful for studying resurgence, and with this data,
we have been able to fix the first few terms in the trans-series.
On the way we observed a very interesting resurgence pattern between
the physical observables, which deserves to be investigated further,
and we hope that our results will spark new research in this field.
We now summarize our method and results.

\subsection{Summary}

We started to investigate the perturbative expansion of the groundstate
energy as the function of the running renormalized coupling. Having
observed factorial growth we switched to the Borel transform and got
insight into the analytical structure by its Pad\'{e} approximant.
We observed a pole singularity at $1$, a cut starting at $2$ and
another cut starting at $-1$, signaling both UV and IR renormalons.
In order to get a more precise analytical continuation of the Borel
transform we applied the conformal mapping method. We then used the
inverse of the Borel transformation by integrating a bit above and
below the real line, so avoiding the singularities. The results had
unwanted imaginary parts, being the complex conjugate of each other.
Since we wanted to understand how precise the real part was we analyzed
the TBA equations directly. As the TBA equation provides an exact
answer, we solved it numerically with very high (30-50 digits) precision.
Comparison with the real part of the inverse Borel transformed revealed
exponentially suppressed corrections. Our aim was to understand these
non-perturbative imaginary and real deviations directly from the perturbative
corrections, thus to establish the first steps into resurgence.

In doing so we used the fact that the leading singularity on the Borel
plane is encoded in the large $n$ behavior of the perturbative coefficients
$\chi_{n}$. Constant asymptotics determines the residue of the pole
singularity, called Stokes constant, while consecutive $1/n$, $1/n(n-1)$,
corrections provides the perturbative expansion of the function multiplying
the logarithmic cut, called the alien derivative of the original functions,
$\Delta_{\omega}\chi$, where $\omega$ is the position of the cut.
The available large number of precise perturbative coefficients enabled
us to determine these coefficients with high (more than 100 digits)
precision. By using simple assumptions on the Stokes constants, (ratios
of powers of $e$ and $\pi$) and the structure of the alien derivatives
(have the same transcendental structure as the original perturbative
coefficients, involving products of odd zeta functions with increasing
transcendentality) enabled us to guess the first $\approx9$ coefficient
exactly and determine the next $60$ with reasonable but decreasing
precisions. The explicit knowledge of the pole term at $1$ and the
leading perturbative coefficient of the first cut starting at $2$
completely agreed with the imaginary ambiguity of the inverse Borel
transform. In order to cancel the ambiguity we had to add these exponentials
to the perturbative series. Since these new non-perturbative terms
are asymptotic by themselves we had to analyze the Borel transform
of the function multiplying the logarithmic cut starting at $2$.
Similar asymptotic analysis revealed a cut starting at $-1$ and another
one starting at $2$. The ambiguity coming from the former one in
the inverse Borel transform was real and its canceling exponential
left a real contribution, which seemed to agree numerically with the
deviation we observed in the comparison with the TBA result. This
was a very reassuring sign, but in order to establish a more precise
matching and to reveal the full trans-series we needed to relate the
various alien derivatives to the original perturbative series. Since
we could not recognize any resurgence structure of the free energy
as the function of the renormalized coupling at this point we started
to analyze the groundstate energy, $\epsilon$ and the density $\rho$
as functions of the TBA variable $B$.

Our asymptotic analysis revealed that $\rho$ has a cut starting at
$-1$ and another one starting at $2$, while $\epsilon$ has the
same structure with an additional pole at $1$. For our big surprise
we managed to relate their alien derivatives at $-1$ with themselves,
i.e. the alien derivative of any of these functions is proportional
to itself multiplied by $\epsilon$. This enabled us to calculate
all consecutive actions of $\Delta_{\pm1}$, which can be expressed
in terms of $\epsilon$ and $\rho$, that is they resurge to themselves.
We called these functions first generation. We could not relate, however,
the new functions appearing by their alien derivatives $\Delta_{2}$
in any way so we called those functions second generation. These second
generation functions resurge also themselves, once we calculated their
$\Delta_{\pm1}$ alien derivatives, but their alien derivatives at
$\Delta_{2}$ are again independent. We called these new functions
the third generation. Our original numerical precision allowed as
to see the structure only up to this point, but already there a beautiful
structure appeared, which definitely deserves a better understanding
and a quantitative description, which might be obtained by analytically
calculating the trans-series expansion directly from the TBA equation.

By using the formula for the alien derivatives of composite functions
we managed to calculate the resurgence properties of the free energy
as a function of the renormalized coupling. These formulas allowed
us to express the $D_{\pm1}$ alien derivatives (in the running coupling)
of the free energy in terms of first generation functions, while the
$D_{2}$ derivatives with first and second generation functions. These
allowed us to calculate the expression $D_{2}D_{2}f$, which contributes
to the leading real deviation from the TBA results. Using the median
resummation prescription based on the alien derivatives $D_{1}f$,
$D_{2}f$ and $D_{2}D_{2}f$ we could reproduce both the imaginary
and real deviations from the TBA result, i.e. we reproduced the real
physical value including non-perturbative exponentially suppressed
terms purely from the perturbative coefficients. These completed the
first steps in the resurgence program and provided the leading terms
in the trans-series ansatz, which we also formulated. Unfortunately
we cannot see yet, how any bridge equation could be derived, which
would relate the functions from different generations to each other.

In summarizing, by determining a large number of high precision perturbative
coefficients in the $O(4)$ model we managed to extract non-perturbative
information and construct the first few terms in the trans-series
exactly. These results are in complete agreement with numerical data
obtained from the conformal mapping method and the direct numerical
solution of the TBA equation.

Our results provide the location of the first few non-trivial saddle
points together with the exact perturbative coefficients coming from
fluctuations around them. It would be very fascinating to confirm
these numbers by direct field theoretic calculations based on the
uniton or other non-perturbative solutions \cite{Cherman:2013yfa,Demulder:2016mja,Dunne:2015eaa,Dunne:2015ywa,Krichever:2020tgp}.
The first non-perturbative saddle is particularly interesting as it
does not have any fluctuation part.

We observed important resurgence properties of the various functions
but these relations showed only the tip of the iceberg. A more systematic
extensive analysis should reveal the full trans-series and their resurgence
structure, i.e. the formulation of the bridge equations. We believe
that our research will spark new activities in this field. As the
$O(4)$ nonlinear sigma model is equivalent to the $SU(2)$ principal
chiral model generalizations for other $N$ are possible in two directions.
In this respect a double scaling limit \cite{Kazakov:2019laa} could
simplify the analysis. Generalizations to other models including those
in \cite{Marino:2019eym,Marino:2019fuy} would be also very interesting.

\subsection{Outline}

The paper is organized as follows: In the next section we summarize
our setup for the $O(4)$ model in a magnetic field and explain the
perturbative calculation of the free energy in the renormalization
group improved running coupling. We also explain how the same perturbative
series can be obtained from the TBA equation. In section \ref{sec:numerical_TBA_etc}
we use the numerically determined high order perturbative coefficients
and analyze the analytical structure of the Borel transform. We also
calculate the inverse Borel transform and compare it to the TBA result,
which we compute numerically. In Section \ref{sec:Resurgence} we
perform the asymptotic analysis of the perturbative coefficients and
quantitatively understand their leading singularity structure, including
their nearest alien derivatives, which we investigate in a similar
fashion. This enables us to fix the first few terms in the trans-series,
but not enough to see any resurgence. We then perform a similar analysis
for the density and the energy density as the functions of $B$, starting
in section \ref{subsec:Resurgence-properties-of-basic-in-B}. Here
we observe a very nice resurgence structure, which we translate to
the original variables in section \ref{sec:Translation-back}, and
summarise in section \ref{sec:Patterns}. We use the various alien
derivatives in section \ref{sec:Median-resummation} to formulate
the median resummation, which agrees with the TBA results and provides
the first few terms in the trans-series. Some technical details are
relegated to two Appendices.

\section{The $O(N)$ model TBA and perturbative coefficients}

\label{sec:TBA_and_coeff}

The 2D $O(N)$ sigma models are exactly soluble, and provide useful
testing grounds for phenomena appearing in QCD. Particles transform
with respect to the fundamental representation of $O(N)$, and scatter
on each other with an integrable elastic scattering matrix, which
is exactly known. In a magnetic field, positively charged particles
condense in the vacuum. These particles scatter diagonally on each
other, and from the thermodynamic limit of the Bethe ansatz an integral
(TBA) equation can be derived for their spectral densities. The systematic
expansion of this integral equation provides a tool to calculate the
perturbative coefficients at very high orders, as we explain in this
section.

\subsection{Perturbative definition of the model}

The 2D $O(N)$ sigma model is an $O(N)$-invariant quantum field theory
of $N$ scalar fields $S_{i}(x,t)$ living on the unit sphere: $\sum_{i=1}^{N}S_{i}^{2}(x,t)=1$.
We are interested in the Euclidean theory in the case when one of
the conserved $O(N)$ charge, say $Q_{12}=\int(S_{1}\partial_{0}S_{2}-S_{2}\partial_{0}S_{1})dx$,
is coupled to a magnetic field: 
\begin{equation}
\mathcal{L}=\frac{1}{2\lambda^{2}}\left\{ \partial_{\mu}S_{i}\partial^{\mu}S_{i}+2ih(S_{1}\partial_{0}S_{2}-S_{2}\partial_{0}S_{1})+h^{2}\left(S_{3}^{2}+\dots+S_{N}^{2}-1\right)\right\} .
\end{equation}
Here $\lambda$ is the bare coupling and the $h$-dependent terms
are chosen such that the Hamiltonian is simply $H(h)=H(0)-hQ_{12}$.
In the perturbative calculations one introduces an infrared regulator
$-2\omega^{2}S_{1}$ (which is put to zero at the end) in the Lagrangian
to fix the ground-state to be $S_{1}=1$ and $S_{i>1}=0$. By expressing
$S_{1}=\sqrt{1-\lambda^{2}(s_{2}^{2}+\dots+s_{N}^{2})}$ with $\lambda s_{i}=S_{i}$
for $i>1$ we can perturbatively expand the free energy 
\begin{equation}
e^{-V\mathcal{F}(h)}=\int\mathcal{D}[s]e^{-\int d^{D}x\mathcal{L}(x)}\quad;\qquad D=2-\varepsilon
\end{equation}
which is regulated in dimensional regularization. The first few terms
in the perturbative expansion read as \cite{Bajnok:2008it} 
\begin{equation}
\mathcal{F}(h)-\mathcal{F}(0)=-\frac{h^{2}}{2\lambda^{2}}+\frac{N-2}{4\pi}h^{2-\varepsilon}\left\{ \frac{1}{\varepsilon}+\frac{\gamma}{2}+\frac{1}{2}\right\} +\lambda^{2}\frac{N-2}{16\pi^{2}}h^{2-2\varepsilon}\left\{ \frac{1}{\varepsilon}+\gamma+\frac{1}{2}\right\} +O(\lambda^{4})
\end{equation}
where $\gamma=\Gamma'(1)+\ln(4\pi)$. UV divergences can be get rid
off by introducing the renormalized coupling $\tilde{g}$, writing $\lambda^{2}=(\mu e^{\frac{\gamma}{2}})^{\varepsilon}Z_{1}\tilde{g}^{2}$
with $Z_{1}=1-\frac{2\beta_{0}\tilde{g}^{2}}{\varepsilon}-\frac{\beta_{1}\tilde{g}^{4}}{\varepsilon}+\frac{4\beta_{0}^{2}\tilde{g}^{4}}{\varepsilon^{2}}\dots$.
The result can be improved by the renormalisation group, which describes
the running of the coupling $\mu\frac{d\tilde{g}}{d\mu}=\beta(\tilde{g})=-\beta_{0}\tilde{g}^{3}-\beta_{1}\tilde{g}^{5}+\dots.$
In the $O(N)$ models $\beta_{0}=\frac{N-2}{4\pi}$ and $\beta_{1}=\frac{N-2}{8\pi^{2}}$.
Higher terms are scheme dependent and in the $\overline{MS}$ scheme
$\beta_{2}^{\overline{MS}}=\frac{(N+2)(N-2)}{64\pi^{3}}$. Thus the
renormalized free energy can be expressed as 
\begin{equation}
\mathcal{F}(h)-\mathcal{F}(0)=-\frac{h^{2}}{2}\left\{ \frac{1}{\tilde{g}^{2}}-2\beta_{0}\left(\ln\frac{\mu}{h}+\frac{1}{2}\right)-2\beta_{1}\tilde{g}^{2}\left(\ln\frac{\mu}{h}+\frac{1}{4}\right)+O(\tilde{g}^{4})\right\} .
\end{equation}
The running of the coupling guaranties that the result is independent
of the renormalization scheme. It is thus natural to introduce a renormalization
group invariant scale $\Lambda=\mu e^{-\int^{\tilde{g}}\frac{dg}{\beta(g)}}=\mu e^{-\frac{1}{2\beta_{0}\tilde{g}^{2}}}\tilde{g}^{-\beta_{1}/\beta_{0}^{2}}[1+\frac{1}{2\beta_{0}}(\frac{\beta_{1}^{2}}{\beta_{0}^{2}}-\frac{\beta_{2}}{\beta_{0}})\tilde{g}^{2}+\dots].$
By taking this scale in the $\overline{MS}$ scheme we can introduce
the running coupling as 
\begin{equation}
\frac{1}{\tilde{\alpha}}+\Delta\ln\tilde{\alpha}=\ln\frac{h}{\Lambda_{\overline{MS}}}\quad;\qquad\Delta=\frac{\beta_{1}}{2\beta_{0}^{2}}=\frac{1}{N-2}.
\end{equation}
The free energy density then has the expansion 
\begin{equation}
\mathcal{F}(h)-\mathcal{F}(0)=-\beta_{0}h^{2}\left\{ \frac{1}{\tilde{\alpha}}-\frac{1}{2}-\frac{\Delta\tilde{\alpha}}{2}+O(\tilde{\alpha}^{2})\right\} .
\end{equation}
It is quite complicated to proceed with the perturbative calculations
at higher orders. Fortunately the theory is integrable and we will
be able to calculate the higher orders systematically. The quantities,
however, on the integrable side are the density $\rho$ and the ground-state
energy density $\epsilon(\rho)$. They are related to the free energy
by Legendre transformation: 
\begin{equation}
\rho=-\frac{\partial\mathcal{F}}{\partial h}\quad;\qquad\epsilon(\rho)=\mathcal{F}(h)-\mathcal{F}(0)+\rho h.
\end{equation}
In order to express the ground-state energy density in terms of the
density one can introduce a new running coupling 
\begin{equation}
\frac{1}{\alpha}+(\Delta-1)\ln\alpha=\ln\frac{\rho}{2\beta_{0}\Lambda_{\overline{MS}}}\label{run1}
\end{equation}
such that 
\begin{equation}
\epsilon(\rho)=\rho^{2}\pi\Delta\left\{ \alpha+\frac{\alpha^{2}}{2}+\Delta\frac{\alpha^{3}}{2}+O(\alpha^{4})\right\} .\label{first3}
\end{equation}
In the following subsections we calculate the higher order terms from
the TBA equation.

\subsection{TBA calculation of the ground-state energy density}

In the infrared description we start with the particle spectrum and
their scattering matrices. The $O(N)$ model has a particle multiplet
which transform in the fundamental (vector) representation of the
$O(N)$ group. This is a relativistic theory and the dispersion relation
can be parametrized in terms of the rapidity as $E(\theta)=m\cosh\theta,p(\theta)=m\sinh\theta$.
The scattering matrix $S_{ij}^{kl}(\theta)$, which depends on the
difference of the rapidities, is non-diagonal and can be calculated
exactly \cite{Zamolodchikov:1977nu}. We are interested in the ground-state
energy in a magnetic field, when the Hamiltonian is modified as $H(h)=H(0)-hQ_{12}$.
If $h>m$ particles of type~$+$ (corresponding to the field $S_{+}=S_{1}+iS_{2}$)
condense into the vacuum. In order to describe this condensate one
can introduce a finite volume $L$ and analyse momentum quantization
via the Bethe Ansatz equation 
\begin{equation}
mL\sinh\theta_{j}-i\sum_{k=1,k\neq j}^{M}\log S(\theta_{j}-\theta_{k})=2\pi n_{j}
\end{equation}
where $S(\theta)$ is the diagonal $S(\theta)=S_{++}^{++}(\theta)$
scattering element 
\begin{equation}
S(\theta)=-\frac{\Gamma(\frac{1}{2}-\frac{i\theta}{2\pi})\Gamma(\Delta-\frac{i\theta}{2\pi})\Gamma(1+\frac{i\theta}{2\pi})\Gamma(\frac{1}{2}+\Delta+\frac{i\theta}{2\pi})}{\Gamma(\frac{1}{2}+\frac{i\theta}{2\pi})\Gamma(\Delta+\frac{i\theta}{2\pi})\Gamma(1-\frac{i\theta}{2\pi})\Gamma(\frac{1}{2}+\Delta-\frac{i\theta}{2\pi})}.
\end{equation}
For $h>m$ the groundstate is filled with particles of rapidities
$\{\theta_{j}\}\in I$ in an interval. Taking the $L\to\infty$ thermodynamic
limit we also have $M\to\infty$ such that the density $\rho=\frac{M}{L}$
is finite. By introducing the rapidity density of states, $\chi(\theta)$
such that $L\chi(\theta)\frac{d\theta}{2\pi}$ is the number of states
in the interval $(\theta,\theta+d\theta)$ the derivative of the BA
equation has a thermodynamic limit: 
\begin{equation}
\chi(\theta)-\int_{-B}^{B}\frac{d\theta'}{2\pi}K(\theta-\theta')\chi(\theta')=m\cosh\theta\label{TBA}
\end{equation}
where $B$ is a function of $h$ and the kernel is 
\begin{equation}
2\pi K(\theta)=-2\pi i\partial_{\theta}\log S(\theta)=\sum_{k=\{\frac{1}{2},\Delta\}}\{\Psi(k+\frac{1}{2}-\frac{i\theta}{2\pi})-\Psi(k+\frac{i\theta}{2\pi})\}+cc
\end{equation}
where $\Psi$ is the digamma function: $\Psi(\theta)=\partial_{\theta}\log\Gamma(\theta$).
The density and energy are obtained simply as 
\begin{equation}
\rho=\int_{-B}^{B}\frac{d\theta}{2\pi}\chi(\theta)\quad;\qquad\epsilon=m\int_{-B}^{B}\frac{d\theta}{2\pi}\cosh\theta\,\chi(\theta).\label{rhoeps}
\end{equation}
These equations depend on $B$, which can be related to the magnetic
field via: $h=\partial_{\rho}\epsilon(\rho)=\frac{\partial\epsilon}{\partial B}/\frac{\partial\rho}{\partial B}$,
following from minimizing $\mathcal{F}=\epsilon-h\rho$ as a function
of $\rho$. Thus either the magnetic field $h$, or the density $\rho$,
or the parameter $B$ can be used as the control parameter. Large
magnetic fields correspond to large densities and large $B$s, thus
the perturbative expansion goes in $1/B$. In the following we explain
our understanding of Volin's perturbative solution of the TBA.

\subsection{Perturbative expansion of the TBA}

The basic idea is to solve the TBA equation in the bulk $\theta\sim0$
and in the edge regions $\theta\sim B$ perturbatively and to match
the two expansions. The calculations are simpler for the resolvent
\begin{equation}
R(\theta)=\int_{-B}^{B}d\theta'\,\frac{\chi(\theta')}{\theta-\theta'}
\end{equation}
which is analytic on the whole complex plane except on the interval
$(-B,B)$ where it has the jump $R(\theta+i0)-R(\theta-i0)=-2\pi i\chi(\theta)$.
The density is obtained from the residue of the resolvent at infinity,
while the energy density through the Laplace transform 
\begin{equation}
\hat{R}(s)=\int_{-i\infty+0}^{i\infty+0}\frac{dz}{2\pi i}e^{sz}R(B+z/2).
\end{equation}
This is related to the Fourier transform of $\chi(\theta)$: 
\begin{equation}
\hat{R}(s)=2e^{-2sB}\tilde{\chi}(2is)\quad;\qquad\tilde{\chi}(\omega)=\int_{-B}^{B}e^{-i\omega\theta}\chi(\theta)d\theta.\label{16star}
\end{equation}
Then, using the $\chi(\theta)=\chi(-\theta)$ symmetry 
\begin{equation}
\frac{\epsilon}{m}=\int_{-B}^{B}\cosh\theta\chi(\theta)\frac{d\theta}{2\pi}=\int_{-B}^{B}e^{\theta}\chi(\theta)\frac{d\theta}{2\pi}=\frac{1}{2\pi}\tilde{\chi}(i)=\frac{e^{B}}{4\pi}\hat{R}(1/2).
\end{equation}

\begin{figure}
\begin{centering}
\includegraphics[width=5cm]{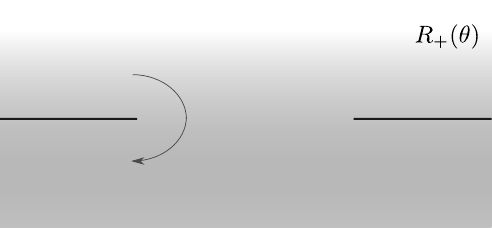}~~~~\includegraphics[width=5cm]{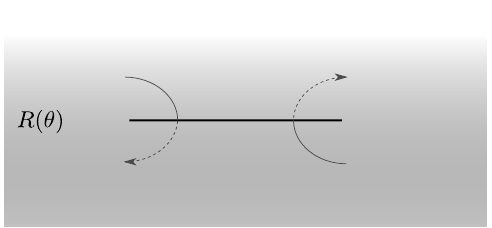}~~~~\includegraphics[width=5cm]{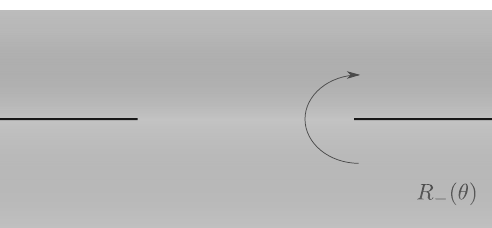}
\par\end{centering}
\caption{Analytic continuation of the resolvent $R(\theta)$ with a short cut
is displayed in the middle. Analytic continuation through the cut
from the upper half plane results in $R_{+}(\theta)$ on the left,
while continuation from the lower half plane results in $R_{-}(\theta)$
displayed on the right. $R_{\pm}(\theta)$ agrees with $R(\theta)$
on the upper/lower half planes and they both have long cuts.}

\label{resolvent}
\end{figure}

In proceeding with the bulk solution we can introduce two functions
$R_{\pm}(\theta)$ obtained by analytical continuations of $R(\theta)$
through the cuts, see Figure \ref{resolvent}. $R_{+}(\theta)$ is
obtained by diving into the cut from above. Thus $R_{+}(\theta)=R(\theta)$
on the upper half plane and has two long cuts $(-\infty,-B)$ and
$(B,\infty)$. $R_{-}(\theta)$ is defined analogously by continuing
from the lower half plane and it also has long cuts. The difference
of the two functions is $R_{+}(\theta)-R_{-}(\theta)=-2\pi i\chi(\theta)$.
By adding the TBA equation at $\theta+i\pi/2$ to that of at $\theta-i\pi/2$
the right hand sides cancel. By taking further into account the $\Psi(z+1)-\Psi(z)=1/z$
property of the digamma function the sum of the TBA equations for
the resolvent takes the form 
\begin{equation}
R_{-}(\theta+\frac{i\pi}{2})-R_{+}(\theta-\frac{i\pi}{2})+R_{+}(\theta+2i\pi\Delta-\frac{i\pi}{2})-R_{-}(\theta-2i\pi\Delta+\frac{i\pi}{2})=0.
\end{equation}
In particular it means that $f(\theta)=R_{+}(\theta+\frac{i\pi}{2}(2\Delta-1))+R_{-}(\theta-\frac{i\pi}{2}(2\Delta-1))$
satisfies $f(\theta)=f(\theta-2i\pi\Delta)$. By making an assumption
for the asymptotic large $B$ behaviour of the resolvent $R(\theta)=F(B)k(\theta/B)$
with $k(x)=\sum_{m=0}^{\infty}k_{m}(x)B^{-m}$ with possible polynomial
$\log B$ dependence in $k_{m}(x)$ one can show that $f(\theta)$
must be a constant. Its antisymmetry then implies that $f(\theta)=0$.

Although it is possible to proceed for generic $O(N)$ models we restrict
our attention to the $O(4)$ model where $\Delta=1/2$, since formulas
are much simpler there. In particular $f(\theta)=0$ implies that
$R_{-}(\theta)=-R_{+}(\theta)$, thus they both are analytic except
for long cuts on the real line. By using the conformal mapping $w=\log\frac{x-1}{x+1}$
the function $R_{+}(w)$ is analytical everywhere, thus must have
the form $k(x)=\sinh\frac{w}{2}\mathcal{F}(\cosh^{2}\frac{w}{2})$.
In the original variable it implies the following expansion 
\begin{equation}
R(\theta)=2\tilde{A}\sqrt{B}\sum_{n,m=0}^{\infty}\frac{c_{n,m}}{B^{m-n}(\theta^{2}-B^{2})^{n+1/2}}.
\end{equation}
where the $c_{n,m}$ coefficients are numerical constants, but the
overall constant $\tilde{A}$ may depend on $B$.The density is obtained
from the residue of the resolvent at infinity 
\begin{equation}
\rho=\tilde{A}\frac{\sqrt{B}}{\pi}\sum_{m=0}^{\infty}c_{0,m}B^{-m}.
\end{equation}
Let us focus on the edge region, where the Wiener-Hopf technique can
be used. Once exponentially small corrections of the form $e^{-B}$
are neglected the Fourier transform of the TBA equation can be separated
for an equation analytic on the upper and another analytic on the
lower half plane \cite{Marino:2019eym}. This determines the Fourier
transform of $\chi$, which is related to $\hat{R}(s)$ by (\ref{16star}).
Using this asymptotic behaviour leads to the following ansatz 
\begin{equation}
\hat{R}(s)=\frac{A}{\sqrt{s}}\frac{\Gamma(1+s)}{\Gamma(\frac{1}{2}+s)}\left(\frac{1}{s+\frac{1}{2}}+\frac{1}{Bs}\sum_{n,m=0}^{\infty}\frac{Q_{n,m}}{B^{n+m}s^{n}}\right)\quad;\qquad A=\frac{me^{B}\sqrt{\pi}}{2\sqrt{2}}
\end{equation}
with constant coefficients $Q_{n,m}$. By re-expanding $R(\theta)$
obtained in the bulk in the edge region, and performing the Laplace
transform, the result can be compared to $\hat{R}(s)$. This gives
$\tilde{A}=A$, and by matching the two parametrizations, all the
unknown coefficients $Q_{n,m}$ and $c_{n,m}$ can be determined.
This was used in \cite{Volin:2009wr,Volin:2010cq} to calculate the
perturbative coefficients for generic $O(N)$ models. The calculations
in the $O(4)$ model take a simpler form, which we present in appendix
\ref{app1}. Having calculated the $Q_{n,m}$ and $c_{n,m}$ coefficients
the energy density and the density can be written as 
\begin{equation}
\epsilon=\frac{me^{B}A}{4\sqrt{2\pi}}\hat{\epsilon}\quad;\qquad\hat{\epsilon}=1+\sum_{k=0}^{\infty}\epsilon_{k}B^{-k-1}\quad;\qquad\epsilon_{k}=\sum_{j=0}^{k}2^{j+1}Q_{j,k-j}
\end{equation}
and 
\begin{equation}
\rho=A\frac{\sqrt{B}}{\pi}\hat{\rho}\quad;\qquad\hat{\rho}=1+\sum_{n=1}^{\infty}c_{0,n}B^{-n}.
\end{equation}
Eventually we are interested in the expansion of $\epsilon/\rho^{2}$
in terms of the running coupling $\alpha$, which is defined by the
following relation 
\begin{equation}
\frac{1}{\alpha}+\frac{1}{2}-B-\frac{1}{2}\log B\alpha=\log2\hat{\rho}.\label{run}
\end{equation}
Here we used the exact $m/\Lambda_{\overline{MS}}$ value \cite{Hasenfratz:1990zz}.
(\ref{run}) can be solved by expanding $B$ in powers of $\alpha$.
The first few terms of this expansion are 
\begin{equation}
B=\frac{1}{\alpha}+\frac{1}{2}-\ln2+\frac{\alpha}{8}+\frac{13-18\,z_{3}}{384}\alpha^{3}+O(\alpha^{4}).
\end{equation}
where $z_{3}=\zeta(3)$ is the zeta function at 3. Using this $B(\alpha)$
relation we can express $\epsilon/\rho^{2}$ in terms of $\alpha$.
As a result we could calculate this expansion analytically at $50$th
order, which goes beyond the result of \cite{Marino:2019eym}.

We then switched to a high precision numerical implementation. This
resulted in 2000 coefficients with 12000 digit precision (for $\rho$
and $\epsilon$). The calculation ran on a PC for 5 days.

\section{Perturbation theory and non-perturbative effects}

\label{sec:numerical_TBA_etc}

Above we described how to expand the TBA to produce results equivalent
to those of ordinary perturbation theory, although allowing thousands
of terms (as powers of $\alpha$ or $1/B$), not just three.

But we can also solve the TBA equation (\ref{TBA}) directly at finite
coupling, without expanding, and find $\rho$ and $\epsilon$ numerically.
These direct results can be compared to those from summing up the
perturbative series, for which we employ Borel resummation methods.
In this section we set up and compare these two techniques, and observe
that the most straightforward resummation omits exponentially small
instanton-like contributions, of order $e^{-8/\alpha}$. These come
multiplied by another power series in $\alpha$, and, thanks to our
very precise numerical results, we are able to fit a few terms.

\subsection{Solving the integral equation}

The TBA equation (\ref{TBA}) is a linear integral equation, which
can be solved analytically for large values of the $B$ by the Wiener-Hopf
method \cite{Hasenfratz:1990zz,Marino:2019eym}.

We are interested in the energy $\epsilon$ and the density $\rho$
as the function of the running coupling $\alpha.$ Rather than work
at fixed $\alpha$, it is much easier to solve the equation at fixed
values of $B$, and then with the help of the formula (\ref{run})
recover points on the functions $\rho(\alpha)$ and $\epsilon(\alpha)$
numerically.

The numerical method we applied was as follows. The unknown $\chi(\theta)$
in the TBA equation (\ref{TBA}) is expanded in even Tschebyshev-polynomials
on $[-B,B]$, up to order $n_{c}-1$ for some odd $n_{c}$: 
\begin{equation}
\chi(\theta)=\sum\limits _{j=1}^{(n_{c}+1)/2}s_{j}\,T_{2j-2}(\theta/B),\qquad T_{n}(x)=\cos(n\arccos x).\label{Tser}
\end{equation}
Inserting this into (\ref{TBA}) and evaluating at the zeros of the
next polynomial $T_{n_{c}}$, namely $\theta_{k}=B\cos[(k-\frac{1}{2})\pi/n_{c}]$
for $k=1,...,n_{c}$, leads to a set of linear algebraic equations
for the coefficients $s_{j}.$ Then formulas in (\ref{rhoeps}) give
$\epsilon$ and $\rho$.

We did this for $B\in\{1,2,...,20\}$, which covers the range from
non-perturbative to highly perturbative. The results can be found
in table \ref{tab:Numerical-TBA-values}. These used $n_{c}=171$,
which is sufficient for 37 digits of precision at $B=20$, and more
at lower values.

\begin{table}
\centering%
\begin{tabular}{c|ccc|c|c|c}
$B$ & $\alpha$ & $\rho$ & $\tilde{\epsilon}_{\mathrm{TBA}}=\epsilon/\rho^{2}$ & $\mbox{prec}_{\mathrm{TBA}}$ & $\tilde{\epsilon}_{\mathrm{conf}}-\tilde{\epsilon}_{\mathrm{TBA}}$ & $\mbox{prec}[\tilde{\epsilon}_{\mathrm{conf}}-\tilde{\epsilon}_{\mathrm{TBA}}${]}\tabularnewline
\hline 
1 & 0.924032 & 0.504839 & 2.35726 & $10^{-142}$ & $-7.26181\cdot10^{-5}$ & $10^{-9}$\tabularnewline
2 & 0.468267 & 2.03329 & 0.975105 & $10^{-106}$ & $-1.16783\cdot10^{-8}$ & $10^{-22}$\tabularnewline
3 & 0.317066 & 6.84148 & 0.596518 & $10^{-87}$ & $-2.57677\cdot10^{-12}$ & $10^{-32}$\tabularnewline
4 & 0.240193 & 21.5688 & 0.430062 & $10^{-75}$ & $-6.43162\cdot10^{-16}$ & $10^{-40}$\tabularnewline
5 & 0.193458 & 65.7062 & 0.336847 & $10^{-68}$ & $-1.71678\cdot10^{-19}$ & $10^{-47}$\tabularnewline
6 & 0.161997 & 195.943 & 0.277064 & $10^{-62}$ & $-4.78057\cdot10^{-23}$ & $10^{-53}$\tabularnewline
7 & 0.139358 & 575.879 & 0.235381 & $10^{-58}$ & $-1.37053\cdot10^{-26}$ & $10^{-58}$\tabularnewline
8 & 0.122281 & 1674.68 & 0.204632 & $10^{-54}$ & $-4.01357\cdot10^{-30}$ & $10^{-54}$\tabularnewline
9 & 0.108938 & 4831.01 & 0.181003 & $10^{-52}$ & $-1.19456\cdot10^{-33}$ & $10^{-52}$\tabularnewline
10 & 0.0982233 & 13848.2 & 0.162274 & $10^{-49}$ & $-3.60099\cdot10^{-37}$ & $10^{-49}$\tabularnewline
11 & 0.0894296 & 39493.9 & 0.147062 & $10^{-47}$ & $-1.09676\cdot10^{-40}$ & $10^{-47}$\tabularnewline
12 & 0.0820823 & 112160. & 0.134460 & $10^{-46}$ & $-3.36936\cdot10^{-44}$ & $10^{-46}$\tabularnewline
13 & 0.0758514 & 317404. & 0.123849 & $10^{-44}$ & $1.81371\cdot10^{-46}$ & $10^{-44}$\tabularnewline
14 & 0.0705003 & 895536. & 0.114792 & $10^{-41}$ & $-5.40468\cdot10^{-45}$ & $10^{-41}$\tabularnewline
15 & 0.0658548 & $2.52018\cdot10^{6}$ & 0.106970 & $10^{-40}$ & $8.50916\cdot10^{-44}$ & $10^{-40}$\tabularnewline
16 & 0.061784 & $7.07623\cdot10^{6}$ & 0.100146 & $10^{-39}$ & $-1.9327\cdot10^{-43}$ & $10^{-39}$\tabularnewline
17 & 0.0581873 & $1.98296\cdot10^{7}$ & 0.0941410 & $10^{-38}$ & $-1.42855\cdot10^{-39}$ & $10^{-38}$\tabularnewline
18 & 0.0549865 & $5.54714\cdot10^{7}$ & 0.0888158 & $10^{-38}$ & $1.57841\cdot10^{-40}$ & $10^{-38}$\tabularnewline
19 & 0.0521196 & $1.54934\cdot10^{8}$ & 0.0840610 & $10^{-37}$ & $2.92388\cdot10^{-40}$ & $10^{-37}$\tabularnewline
20 & 0.0495369 & $4.32133\cdot10^{8}$ & 0.0797896 & $10^{-37}$ & $-9.3026\cdot10^{-39}$ & $10^{-37}$\tabularnewline
\end{tabular}\caption{Some of our results from the numerical solution of the TBA equation
(\ref{TBA}). At fixed $B$, $\epsilon$ and $\rho$ are calculated
from the numerical $\chi(\theta)$, and $\tilde{\epsilon}$, $\alpha$
from them. We also display the numerical uncertainty of the result
together with its deviation from the inverse Borel transformation
based on the conformal mapping method. Note that the entries are truncated
to 6 digits, while our numerical precision is much higher.}
{\footnotesize{} \label{tab:Numerical-TBA-values}}
\end{table}

\subsection{Borel-Padé resummation}

Let us consider the renormalization group improved perturbative series
for the normalized energy density as a function of the running coupling
(\ref{run1}): 
\begin{equation}
\tilde{\epsilon}(\alpha)=\frac{\epsilon}{\rho^{2}}=\frac{\pi}{2}\,\sum\limits _{n=1}^{\infty}\chi_{n}\,\alpha^{n}.\label{epstilde}
\end{equation}
The first few coefficients can be read off from (\ref{first3}) and
a general method \cite{Volin:2009wr,Volin:2010cq} to determine in
principle all coefficients was explained in section \ref{sec:TBA_and_coeff}.
The application of this method allowed us to get 2000 coefficients
of (\ref{epstilde}) with 12000 digits of precision, which makes it
possible to implement the conformal mapping improved Borel resummation
technique with very high precision and compare to the exact TBA data.

This perturbative series has zero radius of convergence, because the
coefficients $\chi_{n}$ grow factorially at large $n$. The Borel
transform is defined by dividing this out, and we choose the following
conventions (removing also some powers of 2): 
\begin{equation}
B(t)=\sum\limits _{n=1}^{\infty}c_{n}\,t^{n},\qquad c_{n}=\frac{\chi_{n+2}\,2^{n+1}}{\Gamma(n+1)}.\label{borelj}
\end{equation}
The inverse of this is a Laplace transform: 
\begin{equation}
\tilde{\epsilon}^{(\pm)}(\alpha)=\frac{\pi}{2}\Bigg[\chi_{1}\,\alpha+\chi_{2}\,\alpha^{2}+\alpha\,\int\limits _{0}^{\infty\pm i0}e^{-\frac{2t}{\alpha}}\,B(t)\,dt\Bigg].\label{epsbor1}
\end{equation}
Applied term-by-term, this would trivially recover (\ref{epstilde}).
But since the Borel series (\ref{borelj}) converges inside the unit
circle, it defines an analytic function. It is the integral of this
function, analytically continued to infinity, which gives a resummation
of $\tilde{\epsilon}(\alpha)$.

\begin{figure}
\begin{centering}
  \includegraphics[width=0.5\textwidth]{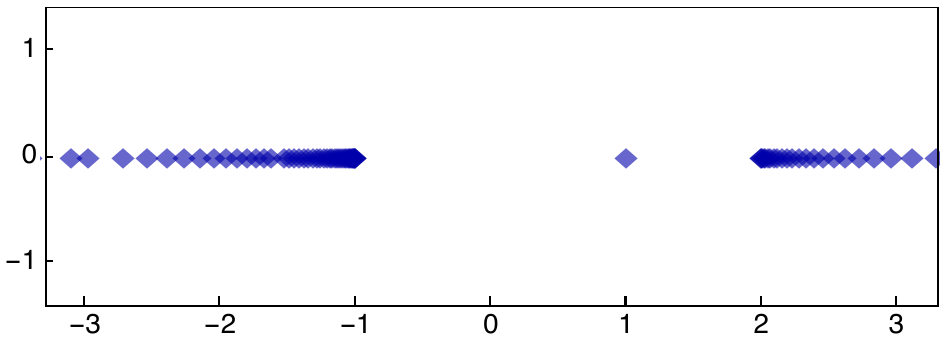}
  \includegraphics[width=0.4\textwidth]{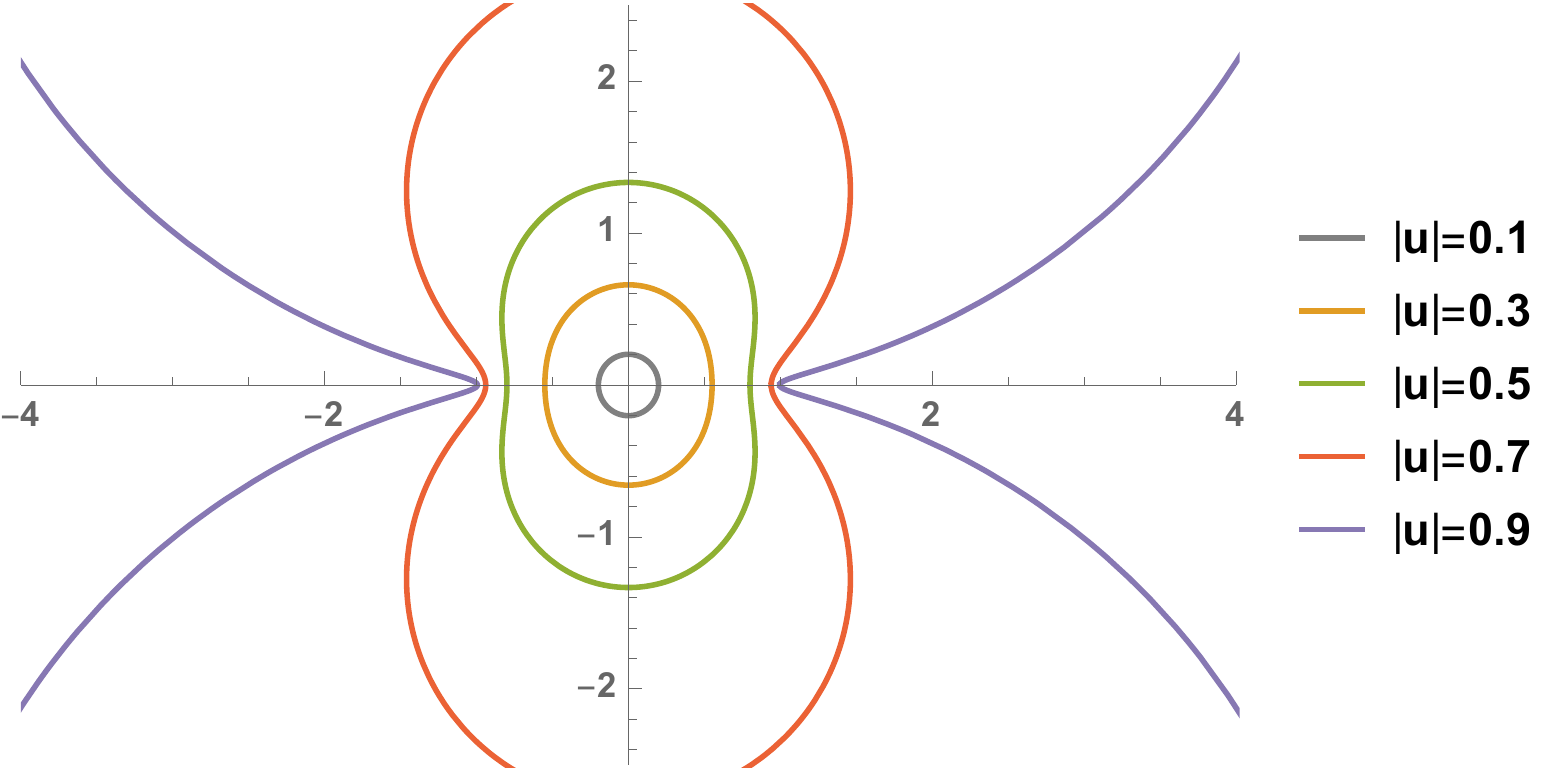}
\par\end{centering}
\caption{Left, positions of the poles of a 100th-order Padé approximant
of the Borel transform of $\hat{\epsilon}$, in the complex $t$ plane. These
accumulate along cuts $t\protect\leq-1$ and $t\protect\geq2$, plus an isolated
pole at $t=1$. 
Right, lines of constant $\left|u\right|$ for the conformal mapping (\ref{ut}),
under which the disk $\left|u\right|<1$ covers the whole $t$ plane,
minus these cuts. \label{fig:Pade-poles-plot}}
\end{figure}

One commonly used representation of the analytic $B(t)$ is given
by Padé approximants. Knowing $N$ coefficients $c_{n}$, we can uniquely
fix the coefficients of this rational function: 
\[
B(t)\approx\frac{\sum_{i=1}^{n}p_{i}t^{i}}{1+\sum_{j=1}^{m}q_{j}t^{j}},\qquad n+m\leq N.
\]
From this, we observe that $B(t)$ has singularities on both the positive
and negative real axes, at $\left|t\right|\geq1$. They are drawn
in figure \ref{fig:Pade-poles-plot}, which shows that there is an
isolated pole singularity at $t=1$ and two dense sets of other poles
which appear to condense into two cuts (starting at $t=-1$ and $t=2$)
as $N$ is increased.

The integration contour must then run either slightly above or below
the positive real axis to avoid the singularities. We indicated this
in (\ref{epsbor1}) by endpoint $\infty\pm i0$, but in practice rotate
the contour off the axis by some small angle. We write $\tilde{\epsilon}^{(\pm)}(\alpha)$
for the results with either choice.

Both $\tilde{\epsilon}^{(+)}$ and $\tilde{\epsilon}^{(-)}$ have
an imaginary part, thus they cannot give the correct physical result.
This non-perturbative ambiguity is defined as 
\begin{equation}
\Delta\tilde{\epsilon}=\frac{1}{2}\left[\tilde{\epsilon}^{(+)}-\tilde{\epsilon}^{(-)}\right]=i{\rm Im}\,\tilde{\epsilon}^{(+)}.
\end{equation}
From our Padé analysis we conclude that the singularities (along the
positive real axis) are: 
\begin{equation}
{\rm a\ pole\ term:}\qquad\frac{\tilde{A}_{0}}{2(1-t)}
\end{equation}
and a logarithmic cut starting at $t=2$. The leading ambiguity is
coming from the pole term: 
\begin{equation}
\frac{i\pi^{2}\alpha\tilde{A}_{0}}{4}{\rm e}^{-\frac{2}{\alpha}}.
\end{equation}
$\tilde{\epsilon}^{(+)}$ and $\tilde{\epsilon}^{(-)}$ share a common
real part, in terms of which we try to approximate the physical result,
and study its deviation from that as a function of the running coupling.

\subsection{Conformal mapping}

For the actual calculation of (\ref{epsbor1}), instead of the Padé
approximants, we used the conformal mapping method. This method is
often applied even to real physical problems \cite{Caprini:2020lff}
to approximate physical quantities better from available perturbative
data. In our case we can make a high precision comparison between
the perturbative and the exact results getting a deeper insight into
the mathematical structure of the deviation.

The Borel-transform (\ref{borelj}) as it stands is not very useful,
since it is still a Taylor-series, which is convergent inside the
unit circle only and thus not applicable in the whole range of the
integration contour. As explained above, the singularities of the
analytically extended $B(t)$ lay on the real axis for $|t|>1$. This
makes it possible to perform a conformal transformation of the form:
\begin{equation}
u(t)=\frac{1-\sqrt{1-t^{2}}}{t}\label{ut}
\end{equation}
which defines a convergent Taylor-series in the unit circle of the
$u$-plane, 
\begin{equation}
\tilde{B}(u)=\sum\limits _{n=1}^{\infty}b_{n}\,u^{n},\qquad|u|<1\label{borelu}
\end{equation}
such that the singularities of $B(t)$ and the integration contours
are mapped to the boundary and into the interior of the unit circle,
respectively. For the actual computations we consider the inverse
transformation: 
\begin{equation}
t(u)=\frac{2\,u}{1+u^{2}}.\label{tu}
\end{equation}
The coefficients $b_{n}$ of the Borel-transform on the $u$-plane
can be determined from those of the original $t$-plane by matching
the small $u$-expansion of the two different representations: 
\begin{equation}
\tilde{B}(u)=\sum\limits _{n=1}^{\infty}b_{n}\,u^{n}=\sum\limits _{n=1}^{\infty}c_{n}\,t(u)^{n}.\label{matchB}
\end{equation}
In this way, we could determine the first 2000 coefficients $b_{n}$.
The improved (approximate) inverse Borel-transform we used in our
calculations is 
\begin{equation}
\tilde{\epsilon}^{(\pm)}(\alpha)\approx\frac{\pi}{2}\Bigg[\chi_{1}\,\alpha+\chi_{2}\,\alpha^{2}+\alpha\,\int\limits _{0}^{\infty\pm i0}dt\;e^{-\frac{2t}{\alpha}}\,\sum\limits _{n=1}^{2000}b_{n}\,u(t)^{n}\Bigg].\label{epsborpmxx}
\end{equation}

\subsection{Fitting the non-perturbative corrections}

We calculated $\tilde{\epsilon}^{(+)}(\alpha)$ as explained above
and compared it to the exact (TBA) results (which are of course real).
We have numerically fitted the correction terms using the ansatz 
\begin{equation}
{\rm Im}\big(\tilde{\epsilon}^{(+)}(\alpha)\big)=\alpha\frac{\pi^{2}\tilde{A}_{0}}{4}{\rm e}^{-2/\alpha}+\alpha{\rm e}^{-4/\alpha}C(\alpha),\qquad C(\alpha)=(c_{1}+c_{2}\alpha+c_{3}\alpha^{2}+\dots)\label{NLOimag}
\end{equation}
\begin{equation}
{\rm Re}\big(\tilde{\epsilon}^{(+)}(\alpha)\big)=\tilde{\epsilon}_{{\rm TBA}}(\alpha)+\alpha{\rm e}^{-8/\alpha}D(\alpha),\qquad D(\alpha)=(d_{1}+d_{2}\alpha+d_{3}\alpha^{2}+\dots).
\end{equation}
Since we know the exact value of the residue of the pole singularity
(see next section) 
\begin{equation}
\tilde{A}_{0}=\frac{16}{{\rm e}\pi}
\end{equation}
our fit provides the NLO term for the imaginary part of the ambiguity.
As our ansatz for the real part shows we found that the leading non-perturbative
ambiguity for the real part is extremely small, it is of fourth order
in the non-perturbative expansion parameter ${\rm e}^{-2/\alpha}$.

We have studied the stability of the first few fit coefficients by
taking higher and higher order fits to the correction functions. This
way we obtained the following quite precise results for $c$ 
\begin{equation}
\begin{split}c_{1} & =-1.70067333(1),\\
c_{2} & =0.637752(1),\\
c_{3} & =-0.1727(1),\\
c_{4} & =-0.027(1),\\
c_{5} & =0.15(3),
\end{split}
\label{ccoffs}
\end{equation}
and $d$ coefficients:
\begin{equation}
\begin{split}d_{1} & =-0.9206(1),\\
d_{2} & =0.575(3),\\
d_{3} & =0.13(3).
\end{split}
\label{dcoffs}
\end{equation}

\section{Resurgence}

\label{sec:Resurgence}

Our data are precise enough to see that, in addition to the simple
Borel resummation, extra non-perturbative corrections are clearly
needed. So far we have only obtained these correction terms by curve
fitting, but our goal is to recover them from the perturbative coefficients
found in section \ref{sec:TBA_and_coeff}. To do this we need resurgence
theory, which this section sets up.

The following sections explore the pattern uncovered, translate back
from an expansion in $1/B$ to an expansion in $\alpha$, and ultimately
use these results to recover the non-perturbative correction terms.

\subsection{Asymptotic coefficients and cuts in the Borel plane}

\label{subB}

\begin{figure}
\begin{centering}
\includegraphics[width=14cm]{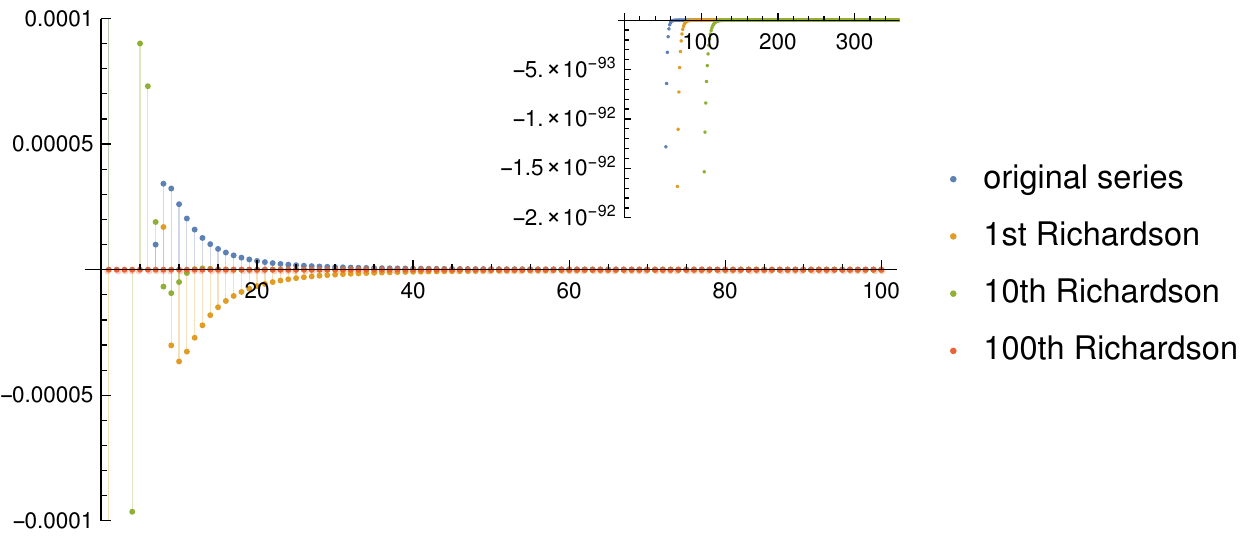}
\par\end{centering}
\caption{Asymptotical investigation of the $a_{n}$ series in (\ref{origi}).
We subtracted the $\frac{16}{\text{e}\pi}$ asymptotical value and
show the first $100$ terms for the original series, the $1^{st}$,
$10^{th}$ and $100^{th}$ Richardson transforms. The inset demonstrates
the deviation in the asymptotics of the $80^{th}$, $90^{th}$ and
$100^{th}$ Richardson transform between $201$ and $600$.}

\label{Richardson}
\end{figure}

The non-perturbative corrections are intimately related to the singularity
structure of the Borel transform $B(t)$. To study these singularities
we use the well-known relation between the cuts (and poles) of $B(t)$
and the asymptotic behaviour of its expansion coefficients.

The series $\{c_{n}\}$ contains two regularly behaving sub-series:
\begin{equation}
a_{n}=c_{2n}+c_{2n-1},\qquad b_{n}=c_{2n}-c_{2n-1},\qquad n=1,2,\dots\label{origi}
\end{equation}
In order to see that we have devided by the correct factorial growth
in (\ref{borelj}) we investigated various Richardson transform of
the $a_{n}$ coefficients. Demonstrative results are presented on
Figure (\ref{Richardson}). By subtracting the leading order behaviour
a similar analysis shows that for asymptotically large $n$ the coefficients
behave as 
\begin{equation}
a_{n}=\tilde{A}_{0}+\frac{\tilde{A}_{1}}{n}+\frac{\tilde{A}_{2}}{n(n-1)}+\sum_{k=3}^{\infty}\frac{\tilde{A}_{k}}{n(n-1)(\cdots)(n-k+1)},
\end{equation}
\begin{equation}
b_{n}=\tilde{B}_{0}+\frac{\tilde{B}_{1}}{n}+\frac{\tilde{B}_{2}}{n(n-1)}+\sum_{k=3}^{\infty}\frac{\tilde{B}_{k}}{n(n-1)(\cdots)(n-k+1)}.\label{origi2}
\end{equation}
Our method to calculate the asymptotic coefficients $\tilde{A}_{k}$,
$\tilde{B}_{k}$ is described in appendix \ref{app2}.

Having computed the asymptotic coefficients we can write down the
singular part of the Borel transform, which is of the form 
\begin{equation}
B^{{\rm sing}}(t)=\frac{\tilde{A}_{0}}{2}\frac{t}{(1-t)}-\frac{\tilde{B}_{0}}{2}\frac{t}{(1+t)}-\frac{1}{2t}\Phi(t)\big[\ln(1-t)+\ln(1+t)\big],\label{singpart}
\end{equation}
where 
\begin{equation}
\Phi(t)=\sum_{k=1}^{\infty}\big[(t+1)\tilde{A}_{k}+(t-1)\tilde{B}_{k}\big]\frac{(t^{2}-1)^{k-1}}{(k-1)!}.
\end{equation}
Around $t=1$ we can expand the coefficient of the log term in powers
of $t-1$ and it becomes 
\begin{equation}
-\frac{\ln(1-t)}{2}\sum_{m=0}^{\infty}p_{m}(t-1)^{m}.\label{tminus1}
\end{equation}
Similarly, around $t=-1$ we have the expansion 
\begin{equation}
=-\frac{\ln(1+t)}{2}\sum_{m=0}^{\infty}q_{m}(-1)^{m}(t+1)^{m}.\label{tplus1}
\end{equation}
The new expansion coefficients $p_{m}$, $q_{m}$ are linear combinations
of the asymptotic coefficients. The first few coefficients are 
\begin{equation}
\begin{split}p_{0} & =2\tilde{A}_{1},\\
p_{1} & =-\tilde{A}_{1}+\tilde{B}_{1}+4\tilde{A}_{2},\\
p_{2} & =\tilde{A}_{1}-\tilde{B}_{1}+2\tilde{B}_{2}+4\tilde{A}_{3},\\
p_{3} & =\tilde{B}_{1}-\tilde{A}_{1}+\tilde{A}_{2}-\tilde{B}_{2}+2\tilde{A}_{3}+2\tilde{B}_{3}+\frac{8}{3}\tilde{A}_{4},
\end{split}
\end{equation}
and 
\begin{equation}
\begin{split}q_{0} & =2\tilde{B}_{1},\\
q_{1} & =-\tilde{B}_{1}+\tilde{A}_{1}+4\tilde{B}_{2},\\
q_{2} & =\tilde{B}_{1}-\tilde{A}_{1}+2\tilde{A}_{2}+4\tilde{B}_{3},\\
q_{3} & =\tilde{A}_{1}-\tilde{B}_{1}+\tilde{B}_{2}-\tilde{A}_{2}+2\tilde{B}_{3}+2\tilde{A}_{3}+\frac{8}{3}\tilde{B}_{4}.
\end{split}
\label{q0}
\end{equation}

\subsection{Results for the asymptotic coefficients}

We observed from our numerical results that all $p_{m}$ coefficients
$m=0,1,\dots$ vanish. We established this fact to more than a hundred
decimal digits for the first few $p_{m}$. Of course the higher coefficients
are less and less precise, but since this observation is in agreement
with our previous finding, namely that apart from a pole at $t=1$
there is no singularity up to the cut starting at $t=2$, from now
on we take it for granted that all $p_{m}$ exactly vanish. We will
also assume that the precision of the non-vanishing coefficients $q_{m}$
is similar to that precision by which the corresponding $p_{m}$ vanish,
that is about 147 digits for $q_{0}$, 142 digits for $q_{1}$, and
gradually decreasing until it remains only 6 digits for $q_{90}$.

Our next observation is that 
\begin{equation}
\tilde{A}_{0}=\frac{16}{{\rm e}\pi}\,(153),\qquad\qquad\tilde{B}_{0}=-\frac{{\rm e}}{4\pi}\,(152).\label{153}
\end{equation}
The numbers in parenthesis indicate that the above relations are satisfied
to 153 (or 152) digits by our estimated results. Clearly we can safely
assume that (\ref{153}) are satisfied exactly.

With the help of the zeta function webpage \texttt{EZ-Face-CECM}\footnote{http://wayback.cecm.sfu.ca/projects/EZFace/}
we were able to find exact expressions for the first few $q_{m}$
coefficients. We found 
\begin{equation}
\begin{split}q_{0} & =0,\\
q_{1} & =\frac{{\rm e}}{4\pi},\\
q_{2} & =\frac{{\rm e}}{4\pi}\left(-\frac{1}{2}-\frac{3}{4}z_{3}\right),\\
q_{3} & =\frac{{\rm e}}{4\pi}\left(\frac{5}{9}-\frac{1}{4}z_{3}\right),\\
q_{4} & =\frac{{\rm e}}{4\pi}\left(-\frac{67}{144}-\frac{3}{64}z_{3}-\frac{135}{256}z_{5}\right),\\
q_{5} & =\frac{{\rm e}}{4\pi}\left(\frac{2797}{7200}-\frac{1}{320}z_{3}-\frac{27}{256}z_{5}-\frac{117}{640}z_{3}^{2}\right),\\
q_{6} & =\frac{{\rm e}}{4\pi}\left(-\frac{4721}{14400}+\frac{239}{11520}z_{3}-\frac{15}{128}z_{5}-\frac{1575}{4096}z_{7}-\frac{47}{1280}z_{3}^{2}\right),\\
q_{7} & =\frac{{\rm e}}{4\pi}\left(\frac{195467}{705600}-\frac{65}{4032}z_{3}+\frac{375}{7168}z_{5}-\frac{225}{4096}z_{7}-\frac{167}{3584}z_{3}^{2}-\frac{3699}{14336}z_{3}z_{5}\right).
\end{split}
\label{q2}
\end{equation}
Here the notation $z_{k}=\zeta(k)={\rm Zeta[k]}$ is used.

\subsection{Alien derivatives}

The notion of alien derivative is a concise and elegant way to characterize
the logarithmic cut (and pole) structure of the Borel transform. We
refer to \cite{Aniceto:2018bis} for details and definition here we
merely summarize the connection between asymptotic coefficients and
alien derivatives.

For a formal asymptotic expansion 
\begin{equation}
\Psi(z)=1+\sum_{n=1}^{\infty}\frac{s_{n}}{z^{n}}
\end{equation}
we introduce the coefficients 
\begin{equation}
c_{n}=\frac{s_{n+1}}{n!}\qquad\qquad n=0,1,\dots
\end{equation}
of its Borel transform and apply the asymptotic analysis explained
in Eqs. (\ref{origi})-(\ref{origi2}) and Eqs. (\ref{singpart})-(\ref{q0}).
We calculate the asymptotic coefficients $\tilde{A}_{m}$, $\tilde{B}_{m}$
and the expansion coefficients $p_{m}$, $q_{m}$. The alien derivative
at $t=1$ is then given by 
\begin{equation}
\Delta_{1}\Psi(z)=-i\pi\left\{ \tilde{A}_{0}+\sum_{m=0}^{\infty}\frac{m!\,p_{m}}{z^{m+1}}\right\} .\label{eq:alien}
\end{equation}
Similarly the alien derivative at $t=-1$ is 
\begin{equation}
\Delta_{-1}\Psi(z)=i\pi\left\{ \tilde{B}_{0}-\sum_{m=0}^{\infty}\frac{m!\,(-1)^{m}q_{m}}{z^{m+1}}\right\} .
\end{equation}

In this language the alien derivative of the energy density at 1 is
a constant and the alien derivative at -1 is characterized by the
coefficients (\ref{q2}), which look like perturbative expansion coefficients
around some saddle point, but we were unable to find any obvious sign
of resurgence in this structure.

\subsection{Results for singularity around $t=2$}

\label{subC}

We have established that there is only a pole singularity at $t=1$
and no cut is starting there. Furthermore the residue of the pole
is exactly known. If we remove this exactly known pole at $t=1$ from
the Borel transform there remains no singularity between $-1$ and
$2$. This allows us to re-expand it around $t=1/2$. The convergence
radius of this new series is $3/2$. We have applied our asymptotic
analysis described above\footnote{One has to rescale the variables by $3/2$ appropriately to absorb
the effect of an increased convergence radius.} to the coefficients of this new series. We find\footnote{For the definition of the alien derivatives $D_{\omega}$ see section
\ref{sec:Translation-back}.} for $f=(2/\pi\alpha)\tilde{\epsilon}$ and expansion parameter $1/x=\alpha/2$
\begin{equation}
D_{2}f=\frac{16i}{{\rm e}^{2}}{\cal F};\qquad\quad{\cal F}=1+\sum_{m=0}^{\infty}\frac{m!\tilde{p}_{m}^{(2)}}{x^{m+1}},
\end{equation}
where the first few coefficients are 
\begin{equation}
\begin{split}\tilde{p}_{0}^{(2)} & =-\frac{3}{4},\\
\tilde{p}_{1}^{(2)} & =\frac{13}{32},\\
\tilde{p}_{2}^{(2)} & =-\frac{99}{256}+\frac{3}{8}z_{3},\\
\tilde{p}_{3}^{(2)} & =-\frac{4655}{36864}-\frac{5}{64}z_{3}.
\end{split}
\label{D2fcoffs}
\end{equation}
$\tilde{p}_{4,5,6}^{(2)}$ are also known analytically. The $q_{m}$
coefficients associated to the singularity at $t=-1$ are already
known (more precisely) from our previous work and can be used to estimate
the precision of the corresponding $\tilde{p}_{m}^{(2)}$ coefficient.
This method suggests that the precision of $\tilde{p}_{m}^{(2)}$
between $m=7$ and $m=51$ gradually decreases from 88 digits to 6
digits. Again, the coefficients (\ref{D2fcoffs}) suggest that they
come from perturbative expansion around some saddle point but we could
not recognize any resurgence here.

The cut characterized by $D_{2}f$ is responsible for the NLO corrections
in (\ref{NLOimag}). Having obtained the overall factor and the first
few coefficients exactly, we can now give exact formulas for the expansion
coefficients $c_{i}$, $i=1,\dots,5$: 
\begin{equation}
\begin{split}c_{1} & =-\frac{4\pi}{{\rm e}^{2}}=-1.7006733263505,\\
c_{2} & =\frac{3\pi}{2{\rm e}^{2}}=0.637752497381,\\
c_{3} & =-\frac{13\pi}{32{\rm e}^{2}}=-0.1727246347,\\
c_{4} & =\frac{3\pi}{4{\rm e}^{2}}\left(\frac{33}{64}-\frac{1}{2}z_{3}\right)=-0.027233132,\\
c_{5} & =\frac{5\pi}{3{\rm e}^{2}}\,\frac{931+576z_{3}}{8192}=0.140424.
\end{split}
\label{exactc}
\end{equation}
The extremely good agreement between the exact coefficients and our
previous fit results (\ref{ccoffs}) (including the error estimates)
makes us confident that our methods are consistent.

\subsection{Resurgence properties of the basic functions $\hat{\epsilon}$ and
$\hat{\rho}$}

\label{subsec:Resurgence-properties-of-basic-in-B}

In this subsection we reconsider the whole resurgence analysis going
back to the more elementary building blocks $\hat{\rho}(B)$ and $\hat{\epsilon}(B)$.
Moreover, we will use the original $B$ parameter to define the new
expansion parameter $1/z$, where $z=2B$.

Using Volin's method, we calculated the expansion coefficients in
the asymptotic series 
\begin{equation}
\hat{\rho}(B)=1+\sum_{n=1}^{\infty}\frac{u_{n}}{B^{n}}=1+\sum_{n=1}^{\infty}\frac{(n-1)!c_{n-1}}{z^{n}},
\end{equation}
where 
\begin{equation}
c_{n}=\frac{2^{n+1}u_{n+1}}{n!}.
\end{equation}
The first two expansion coefficients are 
\begin{equation}
u_{1}=-\frac{3}{8}+\frac{a}{2},\qquad\quad u_{2}=-\frac{15}{128}+\frac{3a}{16}-\frac{a^{2}}{8}.
\end{equation}
The symbol $a$ is again shorthand for the numerical value $a=\ln2$.
We have calculated the first 16 coefficients analytically and observed
that the coefficients depend polynomially on $a$ and that the $u_{p}$
coefficient has order $p$ concerning its zeta-function dependence.
The highest power of $a$ occurring in $u_{p}$ is $a^{p}$ and more
generally this coefficient is a linear combination of terms of the
form $f_{k}a^{k}$, where $f_{k}$ is a zeta-function combination
of order not larger than $p-k$. We calculated the first 2000 coefficients
numerically (putting $a=\ln2$ numerically) with several thousand
digits precision.

The asymptotic analysis of the series 
\begin{equation}
\hat{\epsilon}(B)=1+\sum_{n=1}^{\infty}\frac{\xi_{n}}{B^{n}}
\end{equation}
is completely analogous. The first two coefficients are 
\begin{equation}
\xi_{1}=\frac{1}{4},\qquad\quad\xi_{2}=\frac{9}{32}-\frac{a}{4}
\end{equation}
and by calculating the first 16 coefficients analytically we can establish
that the zeta-function dependence and polynomial $a$-dependence of
$\xi_{p+1}$ is similar to that of $u_{p}$. Remarkably, the $\ln2$-dependence
completely cancels from the density $\tilde{\varepsilon}$, if expressed
in terms of the running coupling $\alpha$.

In this subsection we will use the language of alien derivatives \cite{Aniceto:2018bis}
and will extensively use two of its basic properties:
\begin{itemize}
\item The alien derivative $\Delta_{\omega}$ is a proper derivative, it
is linear and satisfies the Leibniz and chain rules.
\item The pointed alien derivative $\dot{\Delta}_{\omega}={\rm e}^{-\omega z}\Delta_{\omega}$
commutes with the ordinary derivative $\partial/\partial z$.
\end{itemize}
We can go to the Borel plane and study the singularity structure of
the Borel transform 
\begin{equation}
B_{\rho}(t)=\sum_{n=0}^{\infty}c_{n}t^{n}
\end{equation}
and similarly for $B_{\epsilon}(t)$. We use again the method of asymptotic
analysis described in subsection \ref{subB}.

Having computed the asymptotic parameters for both $\hat{\rho}$ and
$\hat{\epsilon}$ very precisely, we were able to recognize the functions
appearing in the $\Delta_{-1}$ alien derivatives. The $\Delta_{\pm1}$
alien derivatives of the two basic functions are given below. 
\begin{equation}
\begin{split}\Delta_{1}\hat{\rho} & =0,\\
\Delta_{1}\hat{\epsilon} & =-4i,
\end{split}
\qquad\qquad\begin{split}\Delta_{-1}\hat{\rho} & =i\hat{\epsilon}\hat{\rho},\\
\Delta_{-1}\hat{\epsilon} & =i\hat{\epsilon}^{2}.
\end{split}
\label{Deltapm1}
\end{equation}
Using these basic derivatives, we can derive further relations for
the basic functions and for the combinations 
\begin{equation}
F=\frac{\hat{\epsilon}}{\hat{\rho}^{2}},\qquad\quad G=\frac{\hat{\epsilon}+\hat{\epsilon}^{\prime}}{\hat{\rho}^{2}}.
\end{equation}
$F$ is useful because it is related to the energy density and $G$
has nice properties as we will see below. We use prime to denote derivative
w.r.t. $z$.

Using (\ref{Deltapm1}) it is easy to verify 
\begin{equation}
\begin{split}\Delta_{1}\Delta_{-1}\hat{\rho} & =4\hat{\rho},\\
\Delta_{1}\Delta_{-1}\hat{\epsilon} & =8\hat{\epsilon},
\end{split}
\qquad\qquad\begin{split}\Delta_{1}\Delta_{-1}F & =\Delta_{-1}\Delta_{1}F=-8F,\\
\Delta_{1}^{2}F & =\Delta_{-1}^{2}F=0.
\end{split}
\end{equation}

It is also easy to see that 
\begin{equation}
\Delta_{1}G=\Delta_{-1}G=0
\end{equation}
and for later purposes we note that 
\begin{equation}
\Delta_{-1}\left(\frac{1}{\hat{\epsilon}}\right)=-i,\qquad\quad\Delta_{1}\left(\frac{1}{\hat{\epsilon}}\right)=\frac{4i}{\hat{\epsilon}^{2}}.
\end{equation}

\subsection{Further alien derivatives}

Since the Borel transform of the function $1/\hat{\epsilon}$ has
only a pole singularity at $t=-1$, after removing this exactly known
pole no singularity remains between $-2$ and $1$. Therefore we re-expanded
the corresponding subtracted Borel transform around $t=-1/2$ and
performed a (rescaled by $3/2$) asymptotic analysis of the coefficients.
We found that 
\begin{equation}
\Delta_{-2}\frac{1}{\hat{\epsilon}}=0,
\end{equation}
implying $\Delta_{-2}\hat{\epsilon}=0$.

The function $G$ was constructed so that 
\begin{equation}
\Delta_{\pm1}G=0,
\end{equation}
hence its expansion around $t=0$ has convergence radius $2$. Applying
the (rescaled by 2) asymptotic analysis we find 
\begin{equation}
\Delta_{-2}G=0.
\end{equation}
This conclusion is based on the observation that $\tilde{B}_{0}=0$,
$q_{0}=0$ to 182 and 177 digits respectively, and $q_{i}=0$ ($i=1,\dots,76$)
to ($172,\dots,6$) digits. At $t=2$ we find 
\begin{equation}
\tilde{A}_{0}=\frac{3}{4\pi},\qquad\quad\tilde{A}_{1}=\frac{3}{4\pi}\left(\frac{9}{4}-2a\right)
\end{equation}
to $182$, $177$ digits respectively, and 
\begin{equation}
\Delta_{2}G=-3iH,\qquad\quad H=1+\sum_{m=0}^{\infty}\frac{m!\tilde{p}_{m}}{(2z)^{m+1}},\qquad\quad p_{m}=\frac{3}{4\pi}\tilde{p}_{m}.
\end{equation}
The first two $\tilde{p}_{m}$ coefficients are 
\begin{equation}
\begin{split}\tilde{p}_{0} & =\frac{9}{2}-4a,\\
\tilde{p}_{1} & =\frac{157}{8}-36a+16a^{2}.
\end{split}
\end{equation}
$\tilde{p}_{2,3,4}$ are also exactly known. The precision of the
exactly known $\tilde{p}_{m}$, $m=0,\dots,4$ is $(177,171,167,162,158)$
respectively and we calculated the $\tilde{p}_{m}$ coefficients up
to $m=76$ numerically. Between $m=1$ and $m=76$ the precision changes
from 171 to 6 digits.

The next observation is that $\tilde{p}_{m}$ is bounded for $m\to\infty$
implying $\Delta_{\pm1}H=0$ and that the Borel transform of $H$
also has convergence radius 2. We note that $\Delta_{-2}G=0$ and
$\Delta_{-2}\hat{\epsilon}=0$ together imply that also 
\begin{equation}
\Delta_{-2}\hat{\rho}=0.
\end{equation}

We performed a (rescaled) asymptotic analysis of the coefficients
of the Borel transform of $H$. We found that 
\begin{equation}
\Delta_{-2}H=0,\qquad\quad\Delta_{2}H=-3i\left\{ 1+\frac{\tilde{p}_{0}}{2z}+\frac{\tilde{p}_{1}}{4z^{2}}+\cdots\right\} ,
\end{equation}
where the first two coefficients are 
\begin{equation}
\tilde{p}_{0}=\frac{11}{2}-4a,\qquad\quad\tilde{p}_{1}=27-44a+16a^{2}.
\end{equation}

Using similar tricks we established that 
\begin{equation}
\begin{split}\Delta_{2}\hat{\rho} & =iR/2,\qquad\quad R=1+\sum_{n=1}^{\infty}\frac{r_{n}}{z^{n}},\\
\Delta_{2}\hat{\epsilon} & =2iE,\qquad\quad E=1+\sum_{n=1}^{\infty}\frac{e_{n}}{z^{n}}.
\end{split}
\label{eq:Delta2rho_eps}
\end{equation}
The first 5 coefficients $r_{1},\dots,r_{5}$ are known analytically:
\begin{equation}
\begin{split}r_{1} & =\frac{1}{2}+a,\\
r_{2} & =-\frac{a}{2}-\frac{a^{2}}{2},\\
r_{3} & =\frac{21}{64}+\frac{3}{4}a^{2}+\frac{a^{3}}{2}+\frac{3}{8}z_{3},\\
r_{4} & =\frac{63}{128}-\frac{105}{64}a-\frac{5}{4}a^{3}-\frac{5}{8}a^{4}-\frac{15}{16}z_{3}-\frac{15}{8}az_{3},\\
r_{5} & =\frac{1485}{512}-\frac{441}{128}a+\frac{735}{128}a^{2}+\frac{35}{16}a^{4}+\frac{7}{8}a^{5}+\frac{105}{16}az_{3}+\frac{105}{16}a^{2}z_{3}+\frac{405}{128}z_{5},
\end{split}
\end{equation}
and similarly $e_{1},\dots,e_{5}$, 
\[
\begin{split}e_{1} & =\frac{1}{4},\\
e_{2} & =\frac{5}{32}-\frac{a}{2},\\
e_{3} & =\frac{57}{128}-\frac{5}{8}a+a^{2},\\
e_{4} & =\frac{2379}{2048}-\frac{171}{64}a+\frac{15}{8}a^{2}-2a^{3}-\frac{27}{32}z_{3},\\
e_{5} & =\frac{41547}{8192}-\frac{2379}{256}a+\frac{171}{16}a^{2}-5a^{3}+4a^{4}-\frac{243}{128}z_{3}+\frac{27}{4}az_{3},
\end{split}
\]
while the rest up to $n=50$ numerically.

The next trick is to consider the function 
\begin{equation}
\eta=\frac{E}{\hat{\rho}^{2}}=1+\sum_{m=0}^{\infty}\frac{m!h_{m}}{(2z)^{m+1}}.
\end{equation}
We found that $\{h_{m}\}$ is bounded for $m\to\infty$ implying $\Delta_{\pm1}\eta=0$.
Thus 
\begin{equation}
\Delta_{1}E=0,\qquad\quad\Delta_{-1}E=2i\hat{\epsilon}E.
\end{equation}
Taking into account that also $\Delta_{\pm1}H=0$, we derive the relations
\begin{equation}
\Delta_{1}R=0,\qquad\quad\Delta_{-1}R=i(4\hat{\rho}E+\hat{\epsilon}R).
\end{equation}
There is also some evidence (a few digits) of the vanishing of $\Delta_{-2}\eta$.
This, combined with $\Delta_{-2}H=0$, implies 
\begin{equation}
\Delta_{-2}R=\Delta_{-2}E=0.
\end{equation}
We have also analysed the $t=2$ singularity. Making the definitions
\begin{equation}
\Delta_{2}E=-\frac{i}{2}\tilde{E},\qquad\quad\Delta_{2}R=-\frac{i}{2}\tilde{R}\label{eq:Delta2E_R}
\end{equation}
the result for the leading expansion terms is 
\begin{align}
\tilde{R} & =1+\left(\frac{1}{4}+a\right)\frac{1}{z}+\left(\frac{5}{32}-\frac{a}{4}-\frac{a^{2}}{2}\right)\frac{1}{z^{2}}+{\rm O}(1/z^{3})\nonumber \\
\tilde{E} & =1+\frac{3}{8z^{2}}+{\rm O}(1/z^{3}).\label{eq:Rtil_Etil_z}
\end{align}

\section{Translation back to the running coupling $\alpha$}

\label{sec:Translation-back} It is natural to use the variable $z=2B$
both for the TBA calculation and for calculation of the coefficients
in Volin's expansion. In perturbation theory on the other hand, the
natural variable is the running coupling $\alpha$. We thus introduce
the expansion parameter 
\begin{equation}
x=\frac{2}{\alpha}.
\end{equation}
The free energy density we were studying originally is given by 
\begin{equation}
\tilde{\epsilon}=\frac{\pi}{z}F(z)=\frac{\pi}{x}f(x)
\end{equation}
and in subsections \ref{subB} and \ref{subC} we studied the resurgence
properties of 
\begin{equation}
f(x)=1+\frac{1}{x}+\frac{1}{x^{2}}+\left(\frac{5}{2}-\frac{3}{4}z_{3}\right)\frac{1}{x^{3}}+\ldots.
\end{equation}
Rewriting (\ref{run}), the relation 
\begin{equation}
z=z(x)
\end{equation}
can be (perturbatively) calculated from 
\begin{equation}
x{\rm e}^{x}=z{\rm e}^{z}\frac{4}{{\rm e}}\hat{\rho}^{2}.\label{plus}
\end{equation}
with the result starting as 
\begin{equation}
z=x+1-2a+\frac{1}{2x}+O(1/x^{3}).
\end{equation}

In the following, we will denote (as before) by $\Delta_{\omega}$
the alien derivative of functions expanded in $1/z$, but use $D_{\omega}$
for alien derivatives of functions expanded in $1/x$. For a composite
function $C=\gamma\circ z$, $C(x)=\gamma(z(x))$ the $D_{\omega}$
alien derivative is given by the formula \cite{composit} 
\begin{equation}
D_{\omega}\gamma(z(x))={\rm e}^{-\omega(z(x)-x)}(\Delta_{\omega}\gamma)(z(x))+\gamma^{\prime}(z(x))(D_{\omega}z)(x).\label{star}
\end{equation}
Applying (\ref{star}) to $\gamma=\hat{\rho}$ and combining it with
the alien derivative of (\ref{plus}) 
\begin{equation}
(1+z)D_{\omega}z+\frac{2z}{\hat{\rho}}D_{\omega}\hat{\rho}=0
\end{equation}
gives 
\begin{equation}
D_{\omega}\gamma=\left(\frac{4z\hat{\rho}^{2}}{x{\rm e}}\right)^{\omega}\left\{ \Delta_{\omega}\gamma-\frac{2x\dot{\gamma}}{(1+x)\hat{\rho}}\Delta_{\omega}\hat{\rho}\right\} \label{starstarstar}
\end{equation}
where dot denotes $d/dx$. We are interested in the alien derivatives
of $f(x)$. Taking 
\begin{equation}
\gamma=\frac{\hat{\epsilon}}{z\hat{\rho}^{2}},\qquad\qquad\left(C=\frac{f(x)}{x}\right)
\end{equation}
we obtain 
\begin{equation}
D_{\omega}f=\left(\frac{4z\hat{\rho}^{2}}{x{\rm e}}\right)^{\omega}\left\{ \frac{f}{\hat{\epsilon}}\Delta_{\omega}\hat{\epsilon}-\frac{2x(f+\dot{f})}{(1+x)\hat{\rho}}\Delta_{\omega}\hat{\rho}\right\} .\label{starstar}
\end{equation}
For $\omega=\pm1$ we get 
\begin{equation}
D_{1}f=\frac{-16i}{{\rm e}}={\rm const.},
\end{equation}
as in subsection \ref{subB}, and 
\begin{equation}
D_{-1}f=\frac{i{\rm e}}{4(1+x)}\left[(1-x)f^{2}-2xf\dot{f}\right].\label{Dminus1f}
\end{equation}
Combining the above two results we also get 
\begin{equation}
D_{1}D_{-1}f=\frac{8}{1+x}\left(f-x\dot{f}\right).
\end{equation}
We have checked that $D_{-1}f$ agrees with the result obtained in
subsection \ref{subB} numerically and, using (\ref{starstar}), we
can verify that 
\begin{equation}
D_{-2}f=0.
\end{equation}

So far we have established that 
\begin{equation}
\begin{split}D_{1}f & =-\frac{16i}{{\rm e}},\\
D_{-1}f & =\frac{i{\rm e}}{4(1+x)}\left\{ (1-x)f^{2}-2xf\dot{f}\right\} ,\\
D_{-2}f & =0,\\
D_{2}f & =\frac{16i}{{\rm e}^{2}}{\cal F},\qquad\quad{\cal F}=\frac{2z\hat{\rho}^{2}}{x}E-\frac{z^{2}\hat{\rho}^{3}(f+\dot{f})}{x(1+x)}R.
\end{split}
\end{equation}
Using (\ref{starstarstar}) $D_{1}{\cal F}=D_{-2}{\cal F}=0$ immediately
follows and after a very long calculation we have verified that 
\begin{equation}
D_{-1}{\cal F}=\frac{i{\rm e}}{2}\left\{ f{\cal F}-\frac{x}{1+x}\frac{{\rm d}}{{\rm d}x}(f{\cal F})\right\} =\frac{i{\rm e}}{2}\left\{ 1+\frac{1}{4x}+\frac{29}{32x^{2}}+\cdots\right\} .
\end{equation}
For the relevant exponantially small correction to TBA,  we also calculated
the expansion of 
\begin{equation}
D_{2}{\cal F}=\frac{16i}{{\rm e}^{2}}\tilde{{\cal F}}=\frac{16i}{{\rm e}^{2}}\left(1-\frac{5}{4x}-\frac{1}{2x^{2}}+\dots\right)
\end{equation}

\section{Resurgence patterns}

\label{sec:Patterns}

Here we collect the results found in the last two sections, to show
clearly the pattern of resurgence. Let us call the functions $\hat{\rho}$,
$\hat{\epsilon}$, $G$, and $f$ elements of the first generation.
Of course, only the first two of these are really fundamental, but
we include $G$ here because it has particularly simple resurgence
structure and $f$ because it is simply related to the free energy
and we were originally interested in it. Similarly, the elements of
the second generation are $R$, $E$, $H$, and ${\cal F}$. Finally,
the third generation consists of $\tilde{R}$, $\tilde{E}$, $\tilde{H}$,
and $\tilde{{\cal F}}$.

\begin{table}
\begin{centering}
\begin{tabular}{cc|ccccccc}
 & Function & $\;-3\;$ & $\;-2\;$ & $-1$ & $\quad0\quad$ & $\;1\;$ & $\;2\;$ & $\;3\;$\tabularnewline
\hline 
1st: & $\hat{\rho}$ & $?$ & $0$ & ${\rm res}$ & $0$ & $0$ & $R$ & $?$\tabularnewline
 & $\hat{\epsilon}$ & $?$ & $0$ & ${\rm res}$ & $0$ & const & $E$ & $?$\tabularnewline
 & $G$ & $+$ & $0$ & $0$ & $0$ & $0$ & $H$ & $?$\tabularnewline
 & $f$ & $?$ & $0$ & ${\rm res}$ & $0$ & const & ${\cal F}$ & $?$\tabularnewline
\hline 
2nd: & $R$ & $?$ & $0$ & ${\rm res}$ & $0$ & $0$ & $\tilde{R}$ & $?$\tabularnewline
 & $E$ & $?$ & $0$ & ${\rm res}$ & $0$ & $0$ & $\tilde{E}$ & $?$\tabularnewline
 & $H$ & $+$ & $0$ & $0$ & $0$ & $0$ & $\tilde{H}$ & $?$\tabularnewline
 & ${\cal F}$ & $?$ & $0$ & ${\rm res}$ & $0$ & $0$ & $\tilde{{\cal F}}$ & $?$\tabularnewline
\hline 
\end{tabular}
\par\end{centering}
\caption{Resurgence structure of first and second generation functions, from
(\ref{eq:summary-first}) to (\ref{eq:summary-last}). Here ``res''
means resurgence to functions encountered above, ``const'' means
a number, and those marked ``+'' are also calculable and nonzero.}

\label{tab:summary}
\end{table}

The resurgence pattern for the elements of the first generation is
\begin{equation}
\begin{split}\Delta_{1}\hat{\rho} & =0,\\
\Delta_{-1}\hat{\rho} & =i\hat{\epsilon}\hat{\rho},
\end{split}
\qquad\quad\begin{split}\Delta_{2}\hat{\rho} & =\frac{i}{2}R,\\
\Delta_{-2}\hat{\rho} & =0,
\end{split}
\label{eq:summary-first}
\end{equation}
\begin{equation}
\begin{split}\Delta_{1}\hat{\epsilon} & =-4i,\\
\Delta_{-1}\hat{\epsilon} & =i\hat{\epsilon}^{2},
\end{split}
\qquad\quad\begin{split}\Delta_{2}\hat{\epsilon} & =2iE,\\
\Delta_{-2}\hat{\epsilon} & =0,
\end{split}
\end{equation}
\begin{equation}
\begin{split}\Delta_{1}G & =0,\\
\Delta_{-1}G & =0,
\end{split}
\qquad\quad\begin{split}\Delta_{2}G & =-3iH,\\
\Delta_{-2}G & =0,
\end{split}
\end{equation}
\begin{equation}
\begin{split}D_{1}f & =\frac{-16i}{{\rm e}},\\
D_{-1}f & =\frac{i{\rm e}}{4(1+x)}\left[(1-x)f^{2}-2xf\dot{f}\right],
\end{split}
\qquad\quad\begin{split}D_{2}f & =\frac{16i}{{\rm e}^{2}}{\cal F},\\
D_{-2}f & =0.
\end{split}
\end{equation}
For the second generation it is 
\begin{equation}
\begin{split}\Delta_{1}R & =0,\\
\Delta_{-1}R & =i[4\hat{\rho}E+\hat{\epsilon}R],
\end{split}
\qquad\quad\begin{split}\Delta_{2}R & =-\frac{i}{2}\tilde{R},\\
\Delta_{-2}R & =0,
\end{split}
\end{equation}
\begin{equation}
\begin{split}\Delta_{1}E & =0,\\
\Delta_{-1}E & =2i\hat{\epsilon}E,
\end{split}
\qquad\quad\begin{split}\Delta_{2}E & =-\frac{i}{2}\tilde{E},\\
\Delta_{-2}E & =0,
\end{split}
\end{equation}
\begin{equation}
\begin{split}\Delta_{1}H & =0,\\
\Delta_{-1}H & =0,
\end{split}
\qquad\quad\begin{split}\Delta_{2}H & =-3i\tilde{H},\\
\Delta_{-2}H & =0,
\end{split}
\end{equation}
\begin{equation}
\begin{split}D_{1}{\cal F} & =0,\\
D_{-1}{\cal F} & =\frac{i{\rm e}}{2}\left(f{\cal F}-\frac{x}{1+x}\frac{{\rm d}}{{\rm d}x}(f{\cal F})\right),
\end{split}
\qquad\quad\begin{split}D_{2}{\cal F} & =\frac{16i}{{\rm e}^{2}}\tilde{{\cal F}},\\
D_{-2}{\cal F} & =0.
\end{split}
\label{eq:summary-last}
\end{equation}

\noindent This patten is illustrated in table \ref{tab:summary}.
In this table the symbol ``const'' means a numerical constant, and
the symbol ``res'' means resurgent, in the sense that the pertinent
alien derivative can be written in terms of functions of the same
or earlier generation. Note that the members of the families $(\hat{\epsilon},E,\tilde{E})$,
$(G,H,\tilde{H})$, and $(f,{\cal F},\tilde{{\cal F}})$ resurge within
the family. Finally, the meaning of the symbol \lq\lq +'' is that
the pertinent alien derivative can be calculated by re-expanding the
Borel transform around $-1/2$. But we do not know how to calculate
$\Delta_{3}$ alien derivatives.

\begin{figure}
\begin{centering}
\includegraphics[width=6cm]{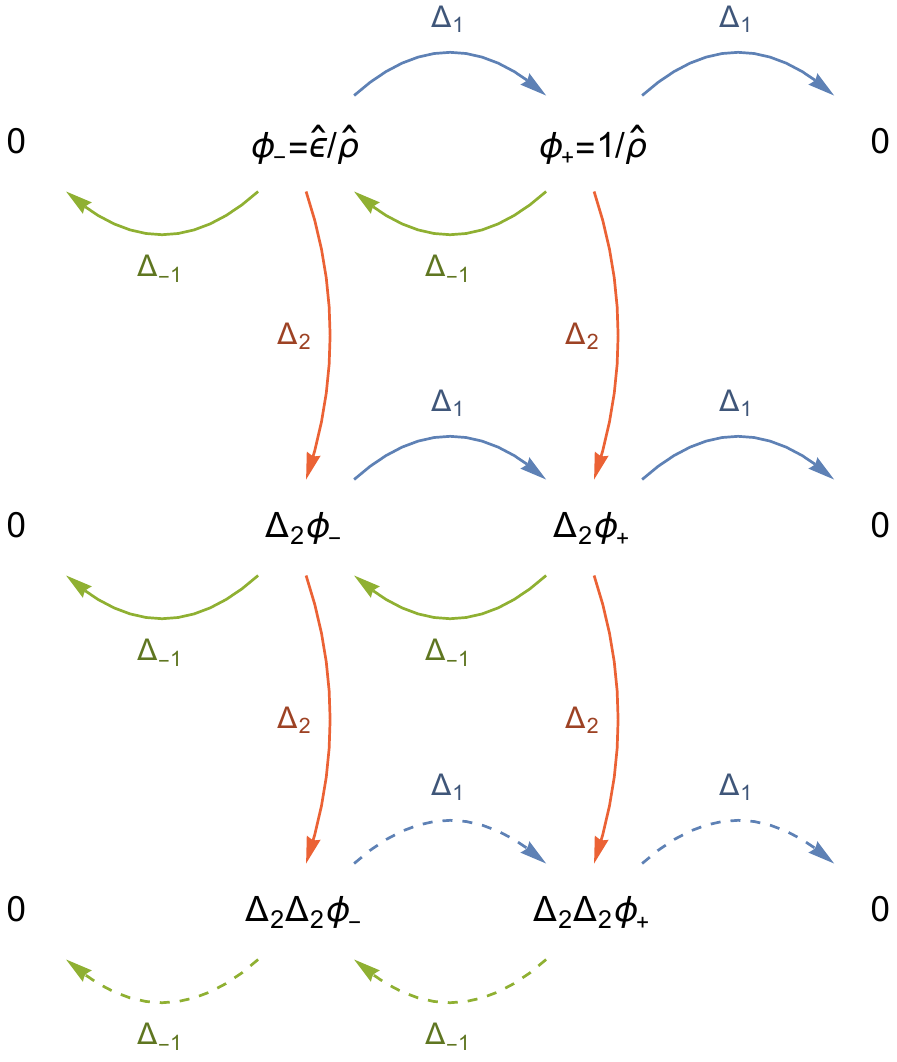}
\par\end{centering}
\caption{Resurgence grid pattern, (\ref{eq:grid-first}) and (\ref{eq:grid-second}),
showing the action of alien derivatives $\Delta_{1}$ (moving right),
$\Delta_{2}$ (moving down), and $\Delta_{-1}$ (moving left). Solid
lines indicate relations derived here from perturbative data, for
the first and second generation functions. Dashed lines indicate a
conjecture that this pattern continues unchanged to the third generation.}

\label{fig:grid}
\end{figure}

In addition to this a pattern of generations, there is another pattern,
which is most easily seen in different variables. Consider the following
representatives of the first generation: 
\begin{equation}
\phi_{-}=\frac{\hat{\epsilon}}{\hat{\rho}}\quad;\qquad\phi_{+}=\frac{1}{\hat{\rho}}.
\end{equation}
These are exchanged by $\Delta_{\pm1}$ as follows:
\begin{equation}
\Delta_{1}\phi_{-}=-4i\phi_{+}\quad;\quad\Delta_{-1}\phi_{+}=-i\phi_{-}\quad;\quad\Delta_{\pm1}\phi_{\pm}=0.
\end{equation}
Acting with $\Delta_{2}$ takes us to the second generation, and we
observe that exactly the same pattern of $\Delta_{\pm1}$ holds:
\begin{equation}
\Delta_{1}(\Delta_{2}\phi_{-})=-4i(\Delta_{2}\phi_{+})\quad;\quad\Delta_{-1}(\Delta_{2}\phi_{+})=-i(\Delta_{2}\phi_{-})\quad;\quad\Delta_{\pm1}(\Delta_{2}\phi_{\pm})=0.\label{eq:grid-first}
\end{equation}
where the basis functions are
\begin{equation}
\Delta_{2}\phi_{-}=2i\frac{E}{\hat{\rho}}-i\frac{\hat{\epsilon}R}{2\hat{\rho}^{2}}\quad;\qquad\Delta_{2}\phi_{+}=-i\frac{R}{2\hat{\rho}^{2}}.
\end{equation}
This pattern is drawn in Figure \ref{fig:grid}, starting with the
first generation at the top.

Acting with $\Delta_{2}$ again will take us to the third generation,
containing also $\tilde{R}$ and $\tilde{E}$. Here we have not derived
the actions of $\Delta_{\pm1}$ above, but the seemingly obvious conjecture
is that the same pattern persists, which would allow us to solve for
these:
\begin{equation}
\Delta_{1}(\Delta_{2}^{2}\phi_{-})=-4i(\Delta_{2}^{2}\phi_{+})\quad;\quad\Delta_{-1}(\Delta_{2}^{2}\phi_{+})=-i(\Delta_{2}^{2}\phi_{-})\quad;\quad\Delta_{\pm1}(\Delta_{2}^{2}\phi_{\pm})=0.\label{eq:grid-second}
\end{equation}

However, this pattern does not capture everything. We have expanded
$\phi_{-}$ on the Borel plane around $t=-3/2$ and performed an asymptotic
analysis, which showed that $\Delta_{-3}\phi_{-}\neq0$. This implies
that the full resurgence pattern must be more complicated.

\section{Median resummation and the cancellation of ambiguities}

\label{sec:Median-resummation}

\begin{figure}
\begin{centering}
\includegraphics[width=9cm]{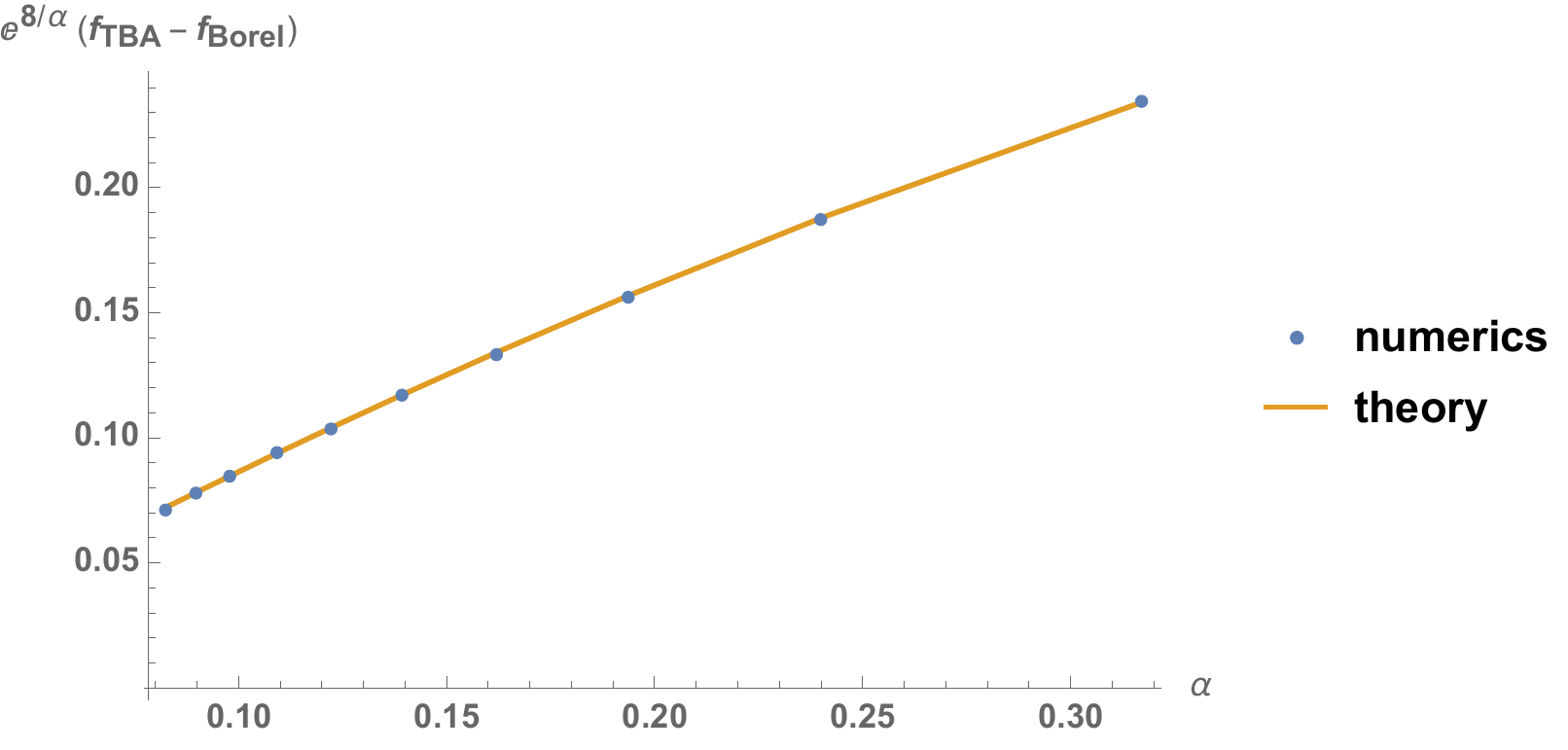}
\par\end{centering}
\caption{Comparison of numerical TBA results (as in Table \ref{tab:Numerical-TBA-values})
to median resummation (\ref{eq:Smed}) assuming $D_{1}D_{3}f=0$.
The lateral Borel resummation $\text{Re}(S_{+}(f))$ has been subtracted
from both, and the exponential prefactor $e^{-8/\alpha}$ has been
divided off. The points shown agree to 3 digits.}

\label{fig:compare}
\end{figure}

Having calculated the relevant alien derivatives of $f=\frac{2}{\alpha\pi}\tilde{\epsilon}$
we are in the position to propose an ambiguity free resummation of
the perturbative series. Clearly the lateral Borel resummations 
\begin{equation}
S_{\pm}(f)=\chi_{1}+\alpha\chi_{2}+\int_{0}^{\infty\pm i0}e^{-tx}B(t)dt\quad;\qquad x=\frac{2}{\alpha}
\end{equation}
are different due the singularities on the positive real line. The
two expressions are related by the Stokes automorphism $\mathfrak{S}$,
which can be written in terms of the alien derivatives as\footnote{Observe that our definition of the alien derivative (\ref{eq:alien})
is such that it is the logarithm of the inverse of the Stokes autormorphism.
This is the opposite which is used in the literature \cite{Dorigoni:2014hea,Aniceto:2013fka}.}
\begin{equation}
S_{+}(f)=S_{-}(\mathfrak{S}f)\quad;\quad S_{-}(f)=S_{+}(\mathfrak{S}^{-1}f)\qquad\mathfrak{S}=\exp\left\{ -\sum_{n=1}^{\infty}e^{-nx}D_{n}\right\} .
\end{equation}
The ambiguity free \emph{median} resummation involves the square root
of the Stokes automorphism and takes the form \cite{Marino:2008ya,Aniceto:2013fka}
\begin{equation}
S_{\mathrm{med}}(f)=S_{-}(\mathcal{\mathfrak{S}}^{\frac{1}{2}}f)=S_{+}(\mathfrak{S}^{-\frac{1}{2}}f)=S_{+}(e^{\frac{1}{2}\sum e^{-nx}D_{n}}f).
\end{equation}
Let us recall the alien derivatives we have calculated 
\begin{equation}
D_{1}f=-\frac{16i}{{\rm e}}\quad;\qquad D_{2}f=\frac{16i}{{\rm e}^{2}}{\cal F}=\frac{16i}{{\rm e}^{2}}\left(1-\frac{3}{4x}+\frac{13}{32x^{2}}-\left(\frac{99}{256}-\frac{3}{8}z_{3}\right)\frac{1}{x^{3}}+\dots\right).
\end{equation}
Since $D_{1}f$ is a constant all higher alien derivatives are vanishing
$D_{k}D_{1}f=0$. We also calculated the first few terms of the alien
derivative of ${\cal F}:$ 
\begin{equation}
D_{1}D_{2}f=D_{1}{\cal F}=0\quad;\qquad D_{2}D_{2}f=\frac{16i}{{\rm e}^{2}}D_{2}{\cal F}=-\left(\frac{16}{{\rm e}^{2}}\right)^{2}\left(1-\frac{5}{4x}-\frac{1}{2x^{2}}+\dots\right).
\end{equation}
Using these results the median resummation takes the form: 
\begin{equation}
S_{\mathrm{med}}(f)=S_{+}(f+\frac{e^{-x}}{2}D_{1}f+\frac{e^{-2x}}{2}D_{2}f+\dots+\frac{e^{-4x}}{8}D_{1}D_{3}f+\frac{e^{-4x}}{8}D_{2}^{2}f+\dots).
\end{equation}
The expression in the bracket provides the ambiguity free trans-series
of the free energy. We did not manage to calculate $D_{3}f$ and higher
derivatives, but already these terms can be compared to the numerically
obtained TBA results. First we can check that the result is real.
Indeed, single alien derivatives are purely imaginary, and the $S_{+}(D_{1}f)$
and $S_{+}(D_{2}f)$ terms cancel the exponentially small imaginary
c-terms (\ref{ccoffs}) coming from $S_{+}(f)$. This can also be
seen from the definition of the Stokes automorphism
\begin{equation}
S_{+}(f)-S_{-}(f)=-S_{+}(e^{-x}D_{1}f+e^{-2x}D_{2}f+\dots)+\dots
\end{equation}
and by noting that the imaginary part of $S_{+}(f)$ is half of the
difference. In calculating the leading real exponential contribution
we point out that the term cancelling the imaginary contribution $S_{+}(D_{2}f)$
also has a real part, which can be read off from 
\begin{equation}
S_{+}(D_{2}f)-S_{-}(D_{2}f)=-S_{+}(e^{-2x}D_{2}D_{2}f)+\dots.
\end{equation}
This combines with the direct $D_{2}^{2}f$ term giving 
\begin{align}
S_{\mathrm{med}}(f) & =\text{Re}(S_{+}(f))-\frac{e^{-4x}}{8}S_{+}(D_{2}^{2}f+D_{1}D_{3}f)+\dots\nonumber \\
 & =\text{Re}(S_{+}(f))+\frac{32}{{\rm e}^{4}}{\rm e}^{-8/\alpha}(1-\frac{5\alpha}{8}-\frac{\alpha^{2}}{8}+\dots)-\frac{e^{-8/\alpha}}{8}S_{+}(D_{1}D_{3}f)+\dots.\label{eq:Smed}
\end{align}
We can compare these with the coefficients we determined previously
\begin{align}
-\frac{2d_{1}}{\pi} & =0.58607 & \frac{32}{{\rm e^{4}}} & =0.58610\nonumber \\
\frac{d_{2}}{d_{1}} & =-0.6246 & -\frac{5}{8} & =-0.625\\
\frac{d_{3}}{d_{1}} & =-0.14 & -\frac{1}{8} & =-0.125\nonumber 
\end{align}
and observe complete agreement within the available precision, which
actually indicates that $D_{1}D_{3}f=0$. This comparison is also
shown in Figure \ref{fig:compare}.

The agreement between the TBA results and the median resummation gives
a strong evidence of the first few terms of the trans-series. The
form of the full trans-series is expected to be 
\begin{equation}
f_{\mathrm{TBA}}=\sum_{m=0}^{\infty}e^{-\frac{2}{\alpha}m}\sum_{n=1}^{\infty}\chi_{n}^{(m)}\alpha^{n-1}
\end{equation}
where $\chi_{n}^{(0)}=\chi_{n}$ are the perturbative coefficients,
$\chi_{n}^{(1)}=-\frac{16i}{{\rm e}}\delta_{n,1}$ are related
to $D_{1}f$, while $\chi_{n}^{(2)}$ to $D_{2}f$.

We have also investigated the relation between the numerical solution
of the TBA equation and the median resummation of the perturbative
series directly for the basic building blocks $\hat{\rho}$ and $\hat{\epsilon}$
in terms of the original TBA variable $B$.

For $\hat{\rho}$ we found by numerically fitting the difference between
the TBA result and the real part of the lateral Borel resummation:
\begin{equation}
\hat{\rho}_{{\rm TBA}}-{\rm Re}[S_{+}(\hat{\rho})]=-\frac{{\rm e}^{-8B}}{32}\left\{ 1+\frac{s_{1}}{B}+\frac{s_{2}}{B^{2}}+\dots\right\} 
\end{equation}
with
\[
s_{1}=0.4717(3),\qquad\qquad s_{2}=-0.061(3)
\]
and the exact guess $1/32$ for the overall coefficient is valid to
5 or 6 digits. The median resummation in this case gives (dropping
the $\Delta_{1}\Delta_{3}\hat{\rho}$ term and using (\ref{eq:Delta2rho_eps}),
(\ref{eq:Delta2E_R}), and (\ref{eq:Rtil_Etil_z}))
\begin{equation}
S_{{\rm med}}(\hat{\rho})-{\rm Re}[S_{+}(\hat{\rho})]=-\frac{{\rm e}^{-8B}}{8}\Delta_{2}\Delta_{2}\hat{\rho}+\dots=-\frac{{\rm e}^{-8B}}{32}\left\{ 1+\frac{\bar{s}_{1}}{B}+\frac{\bar{s}_{2}}{B^{2}}+\dots\right\} 
\end{equation}
where
\[
\bar{s}_{1}=\frac{1}{8}+\frac{a}{2}=0.471574,\qquad\quad\bar{s}_{2}=\frac{5}{128}-\frac{a}{16}-\frac{a^{2}}{8}=-0.0643158.
\]
The good agreement justifies the assumption $\Delta_{1}\Delta_{3}\hat{\rho}=0.$
In the analogous $\hat{\epsilon}$ case we found numerically
\begin{equation}
\hat{\epsilon}_{{\rm TBA}}-{\rm Re}[S_{+}(\hat{\epsilon})]=-\frac{{\rm e}^{-8B}}{8}\left\{ 1+\frac{r_{1}}{B}+\frac{r_{2}}{B^{2}}+\frac{r_{3}}{B^{3}}+\dots\right\} 
\end{equation}
with
\[
r_{1}=0,\qquad\qquad r_{2}=0.0936(16),\qquad\qquad r_{3}=-0.12(3).
\]
This is to be compared to
\begin{equation}
-\frac{1}{8}{\rm e}^{-8B}\Delta_{2}\Delta_{2}\hat{\epsilon}=-\frac{{\rm e}^{-8B}}{8}\left\{ 1+\frac{3}{32B^{2}}+\dots\right\} 
\end{equation}
corresponding to
\[
\bar{r}_{1}=0,\qquad\qquad\bar{r}_{2}=\frac{3}{32}=0.09375.
\]
We do not (yet) have result for $\bar{r}_{3}$. Again, the agreement
suggests $\Delta_{1}\Delta_{3}\hat{\epsilon}=0.$

\subsection*{Acknowledgements}

We thank Ines Aniceto and Daniel Nogradi for useful discussions.

Our work was supported by ELKH, while the infrastructure was provided
by the Hungarian Academy of Sciences. This work was supported in part
by NKFIH grant K134946. M.C.A. was also supported by NKFIH grant FK128789.

\appendix
\bigskip

\section*{Appendix}

\bigskip

\section{Perturbative results for the $O(4)$ model}

\label{app1}

In this appendix we summarize how we solved the recursive equation
for the perturbative coefficients in the $O(4)$ model. The Laplace
transform of the resolvent has the expansion 
\begin{equation}
\hat{R}(s)=\frac{A}{\sqrt{\pi s}}\sum_{m=0}^{\infty}\sum_{n=-m}^{\infty}s^{n}B^{-m}V(n,m)
\end{equation}
where $V(n,m)$ is obtained both from the $Q$ and the $c$ coefficients.
From the $c$ coefficients they can be expressed as 
\begin{equation}
V_{c}(n,m)=\sum_{r=\mathrm{max}(0,-n)}^{m}c_{n+r,m-r}F_{n,r}\quad;\qquad F_{n,r}=2^{1-2r}\frac{\Gamma(\frac{1}{2})\Gamma(\frac{1}{2}-n-r)}{\Gamma(r+1)\Gamma(n+\frac{1}{2})\Gamma(\frac{1}{2}-n-2r)}.
\end{equation}
In the Laplace transform we have to expand in $s$. The explicitly
known terms contribute to $m=0$ 
\begin{equation}
V_{Q}(n,0)=\frac{\Gamma(\frac{1}{2})}{\Gamma(n+1)}\left(\frac{\Gamma(1+x)}{\Gamma(\frac{3}{2}+x)}\right)^{(n)}\quad;\qquad f^{(n)}(x)=\frac{d^{n}f(x)}{dx^{n}}\vert_{x=0}
\end{equation}
while the $Q$ coefficients to $m>0$: 
\begin{equation}
V_{Q}(n,m)=\sum_{r=\mathrm{max}(0,-n-1)}^{m-1}Q_{r,m-r-1}G_{n+r+1}\quad;\qquad G_{n}=\frac{\Gamma(\frac{1}{2})}{\Gamma(n+1)}\left(\frac{\Gamma(1+x)}{\Gamma(\frac{1}{2}+x)}\right)^{(n)}.
\end{equation}
The coefficients, $c_{n,m}$ and $Q_{n,m}$ can be determined by demanding
$V_{Q}(n,m)=V_{c}(n,m)$. In solving these equations it is very natural
to proceed in $m$ and solve all the coefficients iteratively in terms
of smaller $m$ values. For $m=0$ we can start the iteration as $c_{n,0}=V_{Q}(n,0)/F_{n,0}.$
For $m=1$ we can write a separate equation for $n=-1$ giving $Q_{0,0}=c_{0,0}F_{-1,1}/G_{0}$
and another for $n>-1$ resulting in $c_{n,1}=(Q_{0,0}G_{n+1}-c_{n+1,0}F_{n,1})/F_{n,0}$.
Now let us assume that we have already used the equations up to $m-1$
to determine $Q_{i,j}$ for $i+j<m-1$ and $c_{i,j}$ for $j<m$.
We then use the equations for $m$ to determine $Q_{i,j}$ for $i+j=m-1$
and $c_{i,j}$ for $j=m$. In doing so we start with the equation
for $n=-m$ and proceed one by one to $n=-m+1,-m+2,\dots,-1$ to obtain
all the $Q$s: 
\begin{equation}
Q_{-n-1,m+n}=\frac{\sum_{r=-n}^{m}c_{n+r,m-r}F_{n,r}-\sum_{k=-n}^{m-1}Q_{k,m-k-1}G_{n+k+1}}{G_{0}}.
\end{equation}
Finally we use the equation for $n>-1$ to get the $c$s: 
\begin{equation}
c_{n,m}=\frac{\sum_{k=0}^{m-1}Q_{k,m-k-1}G_{n+k+1}-\sum_{r=1}^{m}c_{n+r,m-r}F_{n,r}}{F_{n,0}}.
\end{equation}
We implemented this iterative calculation in Mathematica. In order
to speed up the calculation we use the $\Gamma$-function identity
\begin{equation}
\frac{\Gamma(\frac{1}{2})\Gamma(1+x)}{\Gamma(\frac{1}{2}+x)}=\frac{x\Gamma(x)^{2}}{\Gamma(2x)}2^{2x-1}=\frac{\Gamma(1+x)^{2}}{\Gamma(1+2x)}2^{2x}
\end{equation}
thus we have to expand the $\Gamma$-function only around $1$. We
then used the functional relation $\Gamma(x)\Gamma(1-x)=\frac{\pi}{\sin(\pi x)}$
together with the following recursive expression 
\begin{equation}
\frac{1}{\Gamma(x)}=\sum_{k=1}^{\infty}a_{k}x^{k}\quad;\qquad a_{n}=na_{1}a_{n}-a_{2}a_{n-1}+\sum_{k=2}^{n}(-1)^{k}\zeta(k)a_{n-k}
\end{equation}
to speed up the calculation.

\section{Asymptotic coefficients}

\label{app2}

In this appendix we summarize our asymptotic analysis.

\subsection{Transformation of the coefficient series}

Let us denote the series $\{x_{1},x_{2},\dots,x_{n},\dots\}$ symbolically
by $\xi$. $\xi$ can also be represented by the series of asymptotic
coefficients $\Xi=\{X_{0},X_{1},\dots,X_{k},\dots\}$, where 
\begin{equation}
x_{n}=X_{0}+\frac{X_{1}}{n}+\frac{X_{2}}{n(n-1)}+\sum_{k=3}^{\infty}\frac{X_{k}}{n(n-1)(\cdots)(n-k+1)}.\label{xn}
\end{equation}
Let us define the transformation 
\begin{equation}
T\xi=\xi^{\prime}=\{x_{1}^{\prime},x_{2}^{\prime},\dots,x_{n}^{\prime},\dots\},
\end{equation}
where 
\begin{equation}
x_{n}^{\prime}=(n+1)(n+2)[x_{n+1}-x_{n+2}],\qquad n=1,2,\dots.
\end{equation}
In terms of the asymptotic coefficients we have 
\begin{equation}
T\Xi=\Xi^{\prime}=\{X_{0}^{\prime},X_{1}^{\prime},\dots,X_{k}^{\prime},\dots\},
\end{equation}
where 
\begin{equation}
X_{k}^{\prime}=(k+1)X_{k+1},\qquad k=0,1,\dots.
\end{equation}
Let us now define 
\begin{equation}
\xi^{(j)}=\frac{1}{j!}T^{j}\xi.
\end{equation}
This can be calculated, starting from $\xi^{(0)}=\xi$, using the
recursive formula 
\begin{equation}
\xi^{(j)}=\frac{1}{j}T\xi^{(j-1)}.
\end{equation}
For the asymptotic coefficients we get 
\begin{equation}
\Xi^{(j)}=\{X_{0}^{(j)},X_{1}^{(j)},\dots,X_{k}^{(j)},\dots\},
\end{equation}
with 
\begin{equation}
X_{k}^{(j)}=\binom{k+j}{j}X_{k+j}.
\end{equation}
Most importantly, 
\begin{equation}
X_{0}^{(j)}=X_{j}.
\end{equation}
After $j$ steps the $j^{{\rm th}}$ coefficient of the original series
is promoted to leading term.

\subsection{Calculation of the constant term}

Let us study the series of transformations 
\begin{equation}
\{x_{n}^{[0]}\}=\{x_{n}\}\Longrightarrow\{x_{n}^{[1]}\}\Longrightarrow\{x_{n}^{[2]}\}\Longrightarrow\cdots,
\end{equation}
where 
\begin{equation}
x_{n}^{[s]}=\frac{1}{s}\left[(n+1-s)x_{n}^{[s-1]}-(n+1-2s)x_{n-1}^{[s-1]}\right],\quad n=s+1,s+2,\dots.
\end{equation}
After the first step, the $1/n$ term is eliminated and the asymptotic
expansion becomes 
\begin{equation}
x_{n}^{[1]}=X_{0}+\sum_{k=2}^{\infty}\frac{(1-k)X_{k}}{(n-1)(\cdots)(n-k)}.
\end{equation}
In the subsequent transformations, the O$(1/n^{2})$, O$(1/n^{3})$,
etc. terms are eliminated step by step, and after $s$ steps the asymptotic
form is 
\begin{equation}
x_{n}^{[s]}=X_{0}+\sum_{k=s+1}^{\infty}\frac{(1-k)(2-k)(\cdots)(s-k)X_{k}}{(n-s)(n-s-1)(\cdots)(n-s+1-k)},\label{fact}
\end{equation}
i.e. 
\begin{equation}
x_{n}^{[s]}=X_{0}+{\rm O}(1/n^{s+1}).
\end{equation}
The $s^{{\rm th}}$ approximant of the leading (constant) term $X_{0}$
is thus 
\begin{equation}
X_{0}[s]=x_{{\rm max}}^{[s]},\label{max}
\end{equation}
where $x_{{\rm max}}^{[s]}$ is the last available member of the series
(max is common to all $s$). Applying (\ref{max}) for the transformed
series $\xi^{(m)}$ we get an approximation for $X_{m}$ in the form
\begin{equation}
X_{m}[s]=x_{{\rm max}}^{(m)[s]}.
\end{equation}

\section{Alien derivative of function composition}

\label{sec:composition-appendix}

In this appendix, we prove the relation which we used for the alien
derivative of composite resurgent function $\psi[\phi(x)]$, where
both $\psi$ and $\phi$ are resurgent functions of $x$. This relation
was proposed in \cite{composit}. Here, we present its proof. We start
by the following form of the composite function 
\begin{equation}
\psi[\phi(x)]=\sum_{n=0}^{\infty}\frac{1}{n!}\left(\frac{d^{n}}{dx^{n}}\psi(x)\right)\Big(x\,\varepsilon(x)\Big)^{n}\,,\label{composition}
\end{equation}
which is suggested by the theorem 0.3.2. of \cite{composit} and the
following form is assumed for the resurgent function $\phi(x)$ 
\begin{equation}
\phi(x)=x\,\left(1+\varepsilon(x)\right)\,.
\end{equation}

The alien derivative of the composite function (\ref{composition}),
has two terms 
\begin{equation}
\Delta_{\omega}\psi[\phi(x)]=\sum_{n=0}^{\infty}\frac{1}{n!}\left\{ \left(\Delta_{\omega}\frac{d^{n}}{dx^{n}}\psi(x)\right)\Big(x\,\varepsilon(x)\Big)^{n}+\left(\frac{d^{n}}{dx^{n}}\psi(x)\right)\Delta_{\omega}\Big(x\,\varepsilon(x)\Big)^{n}\right\} \,,\label{Delta-composed}
\end{equation}
where, we used the Leibniz rule for $\Delta_{\omega}$.

In the first term of (\ref{Delta-composed}), we should commute the
alien derivative with the $n$-th ordinary derivative. We know that
the ordinary and alien derivative do not commute, 
\begin{equation}
[\Delta_{\omega},\frac{d}{dx}]\phi(x)=-\,\omega\Delta_{\omega}\phi(x)\,.
\end{equation}
Some algebraic calculation shows that 
\begin{equation}
[\Delta_{\omega},\frac{d^{n}}{dx^{n}}]\phi(x)=\sum_{\ell=0}^{n-1}\frac{n!}{\ell!\,(n-\ell)!}(-\omega)^{n-\ell}\,\frac{d^{\ell}}{dx^{\ell}}\Delta_{\omega}\phi(x)\,.\label{n-th}
\end{equation}

Using this result, we change the order of $\Delta_{w}$ and $\frac{d^{n}}{dx^{n}}$
in the first term of (\ref{Delta-composed}), 
\begin{eqnarray}
\sum_{n=0}^{\infty}\frac{1}{n!}\left(\Delta_{\omega}\frac{d^{n}}{dx^{n}}\psi(x)\right)\Big(x\,\varepsilon(x)\Big)^{n} & = & \sum_{n=0}^{\infty}\sum_{\ell=0}^{n}\frac{(-\omega)^{n-\ell}}{\ell!\,(n-\ell)!}\,\left(\frac{d^{\ell}}{dx^{\ell}}\Delta_{\omega}\psi(x)\right)\Big(x\,\varepsilon(x)\Big)^{n}\nonumber \\
 & = & \sum_{\ell=0}^{\infty}\sum_{n=\ell}^{\infty}\frac{(-\omega)^{n-\ell}}{\ell!\,(n-\ell)!}\,\left(\frac{d^{\ell}}{dx^{\ell}}\Delta_{\omega}\psi(x)\right)\Big(x\,\varepsilon(x)\Big)^{(n-\ell)+\ell}\nonumber \\
 & = & \left[\sum_{k=0}^{\infty}\frac{(-\omega)^{k}}{k!}\Big(x\,\varepsilon(x)\Big)^{k}\right]\sum_{\ell=0}^{\infty}\left(\frac{d^{\ell}}{dx^{\ell}}\Delta_{\omega}\psi(x)\right)\frac{\Big(x\,\varepsilon(x)\Big)^{\ell}}{\ell!}\nonumber \\
 & = & \exp{\left[-\omega\,x\,\varepsilon(x)\right]}\cdot\left(\Delta_{\omega}\psi\right)[\phi(x)]\,.
\end{eqnarray}
where, we changed the order of the summation on $n$ and $\ell$ in
the second line. In the third line, we changed the variable from $n$
to $k=n-\ell$ which runs from zero to infinity. In the last line,
we used (\ref{composition}) to write the closed form of the second
bracket of the third line, i.e. $\left(\Delta_{\omega}\psi\right)[\phi(x)]$.

The second term of (\ref{Delta-composed}) is 
\begin{eqnarray}
\sum_{n=0}^{\infty}\frac{1}{n!}\left(\frac{d^{n}}{dx^{n}}\psi(x)\right)\Delta_{\omega}\Big(x\,\varepsilon(x)\Big)^{n} & = & \Big(\Delta_{\omega}\phi(x)\Big)\cdot\sum_{n=0}^{\infty}\frac{1}{(n-1)!}\left(\frac{d^{n-1}}{dx^{n-1}}\psi'(x)\right)\Big(x\,\varepsilon(x)\Big)^{n-1}\,,\nonumber \\
 & = & \Big(\Delta_{\omega}\phi(x)\Big)\cdot\psi'[\phi(x)]\,.
\end{eqnarray}

Combining the two parts of (\ref{Delta-composed}) gives the desired
result for the alien derivative of the composite function 
\begin{equation}
\Delta_{\omega}\psi[\phi(x)]=\exp{\left[-\omega\,x\,\varepsilon(x)\right]}\cdot\left(\Delta_{\omega}\psi\right)[\phi(x)]+\Big(\Delta_{\omega}\phi(x)\Big)\cdot\psi'[\phi(x)]\,.
\end{equation}

\bibliographystyle{elsarticle-num}
\bibliography{borel}

\end{document}